\RequirePackage[2024/06/01]{latexrelease}
\documentclass[
 reprint,
 amsmath,amssymb,
 aps,
 superscriptaddress,
]{revtex4-2}

\usepackage{array}[=2016-10-06]
\preprint{APS/123-QED}
\usepackage{braket}
\usepackage[utf8]{inputenc}
\usepackage[T1]{fontenc}
\usepackage[normalem]{ulem}
\usepackage{amsmath}
\usepackage{amsthm}
\usepackage{amssymb}
\usepackage{subfiles} 
\usepackage{comment}
\usepackage{graphicx}
\usepackage[normalem]{ulem}
\usepackage{makecell}
\usepackage{cancel}
\setcellgapes{3.5pt}
\usepackage{bm,dcolumn,amsmath,graphicx,amsfonts,amssymb}
\usepackage[utf8]{inputenc}
\usepackage[dvipsnames]{xcolor}
\usepackage{booktabs} 
\usepackage[dvipsnames]{xcolor}
\usepackage{graphicx}
\usepackage[colorlinks=true, allcolors=blue]{hyperref}
\makegapedcells
\usepackage{orcidlink}
\usepackage{array}
\usepackage{booktabs}

\newcommand{\eref}[1]{Eq.~(\ref{#1})}
\newcommand{\tref}[1]{Table~\ref{#1}}

\begin{document}
\title{Shielded inner-shell transitions in atomic samarium for tests of fundamental physics}

\author{R. Aramyan\,\orcidlink{0009-0007-2773-7343}}
 \email[Contact author: ] {aramyanr@uni-mainz.de}
\affiliation{
 Johannes Gutenberg-Universit\"at Mainz, 55128 Mainz, Germany
}
\affiliation{
Helmholtz Institute Mainz, 55099 Mainz, Germany
}
\author{D. Budker\,\orcidlink{0000-0002-7356-4814}}
\affiliation{
 Johannes Gutenberg-Universit\"at Mainz, 55128 Mainz, Germany
}
\affiliation{
Helmholtz Institute Mainz, 55099 Mainz, Germany
}
\affiliation{GSI Helmholtzzentrum f\"ur Schwerionenforschung GmbH, 64291 Darmstadt, Germany}
\affiliation{
 Department of Physics, University of California, Berkeley, CA 94720, USA
}
\author{V.~A. Dzuba\,\orcidlink{0000-0003-2758-5574}}
\affiliation{
  School of Physics, University of New South Wales, Sydney 2052, Australia
}
\author{V.~V. Flambaum\,\orcidlink{0000-0001-8643-7374}}
\affiliation{
  School of Physics, University of New South Wales, Sydney 2052, Australia
}
\author{S.~G. Porsev\,\orcidlink{0000-0003-0417-2726}}
\affiliation{Department of Physics and Astronomy, University of Delaware, Newark, Delaware 19716, USA}
\author{M.~S. Safronova\,\orcidlink{0000-0002-1305-4011}}
\affiliation{Department of Physics and Astronomy, University of Delaware, Newark, Delaware 19716, USA}
\author{O. Tretiak\,\orcidlink{0000-0002-7667-2933}}
\affiliation{
 Johannes Gutenberg-Universit\"at Mainz, 55128 Mainz, Germany
}
\affiliation{
Helmholtz Institute Mainz, 55099 Mainz, Germany
}
\author{K. Zhang\,\orcidlink{0000-0003-4370-7877}}
 \email[Contact author: ] {kzhang@uni-mainz.de}
\affiliation{
 Johannes Gutenberg-Universit\"at Mainz, 55128 Mainz, Germany
}
\affiliation{
Helmholtz Institute Mainz, 55099 Mainz, Germany
}

\begin{abstract}
Forbidden atomic transitions provide some of the most stringent
low-energy tests of physics beyond the Standard Model, with sensitivity
set by the interplay between the sought-for signals and
systematics suppressed by symmetry. Here we identify the previously
unobserved $4f^{6}6s^{2}\,{}^{5}$D$_{0}$ level of neutral samarium at
$14\,564.90(2)\,\mathrm{cm}^{-1}$, opening the
${}^{7}$F$_{0}\rightarrow{}^{5}$D$_{0}$ inner-shell transition for
precision spectroscopy. Candidate lines extracted from dual-comb
absorption spectra were assigned using double-resonance
population-depletion and sequential-excitation measurements.  The observed pressure broadening, $0.12(2)\,\mathrm{MHz/torr}$, and pressure shift, $-0.145(4)\,\mathrm{MHz/torr}$, indicate an inner-shell $4f$-transition shielded from external perturbations. Many-body calculations predict a metastable lifetime in the range $120$--$200\,\mathrm{ms}$, corresponding to a quality factor of order $\mathcal{Q}\sim10^{14}$,
large sensitivity coefficients for variation of the fine-structure constant,
and a nuclear-spin-dependent parity-violation amplitude comparable to that
of cesium. Crucially, the $J=0\rightarrow J=0$ selection rule suppresses by symmetry 
both the nuclear-spin-independent parity-violation channel and the M1
and E2 backgrounds that complicated previous heavy-atom experiments,
yielding a uniquely clean window onto the nuclear anapole moment.
The two stable spin-$7/2$ isotopes of samarium provide a remarkable opportunity to largely cancel atomic-structure uncertainties by measuring the ratio of parity-violation effects in the two isotopes.
These results establish neutral samarium as a platform for inner-shell
precision spectroscopy and tests of physics beyond the Standard Model.
\end{abstract}

\maketitle

\textit{Introduction:} Fundamental but subtle atomic effects, such as parity violation (PV), manifest themselves as a contribution to forbidden transition amplitudes. To study such effects, one seeks atomic systems where the sought-for effects are enhanced, while parasitic systematic effects are suppressed. Here, we present such a system in samarium as a platform for searching for beyond-standard-model physics.

Extensive studies of samarium spectra were carried out in the past. Early on, traditional absorption and emission spectroscopy were employed to obtain data on low-lying states of the atom (see the discussion and references in \cite{Martin|1978|Energy_levels,Barkov1988}). With the development of laser spectroscopy, the focus switched to the studies of high-lying even- \cite{Pulhani2005,Li2011,Gomonai2012,Shah2014,Sahoo2020} and odd-parity states \cite{Feng2008,LiMing2011}, autoionizing states \cite{Sufen1989,WenJie2009,Qin2010,QinWenJie2010}, Rydberg states \cite{Jayasekharan2000}, and the ionization potential \cite{klaus2014,Worden1978}. However, the available spectroscopic data on atomic samarium are far from complete.

\begin{figure}[t]
	\centering
	\includegraphics[width=0.9\linewidth]{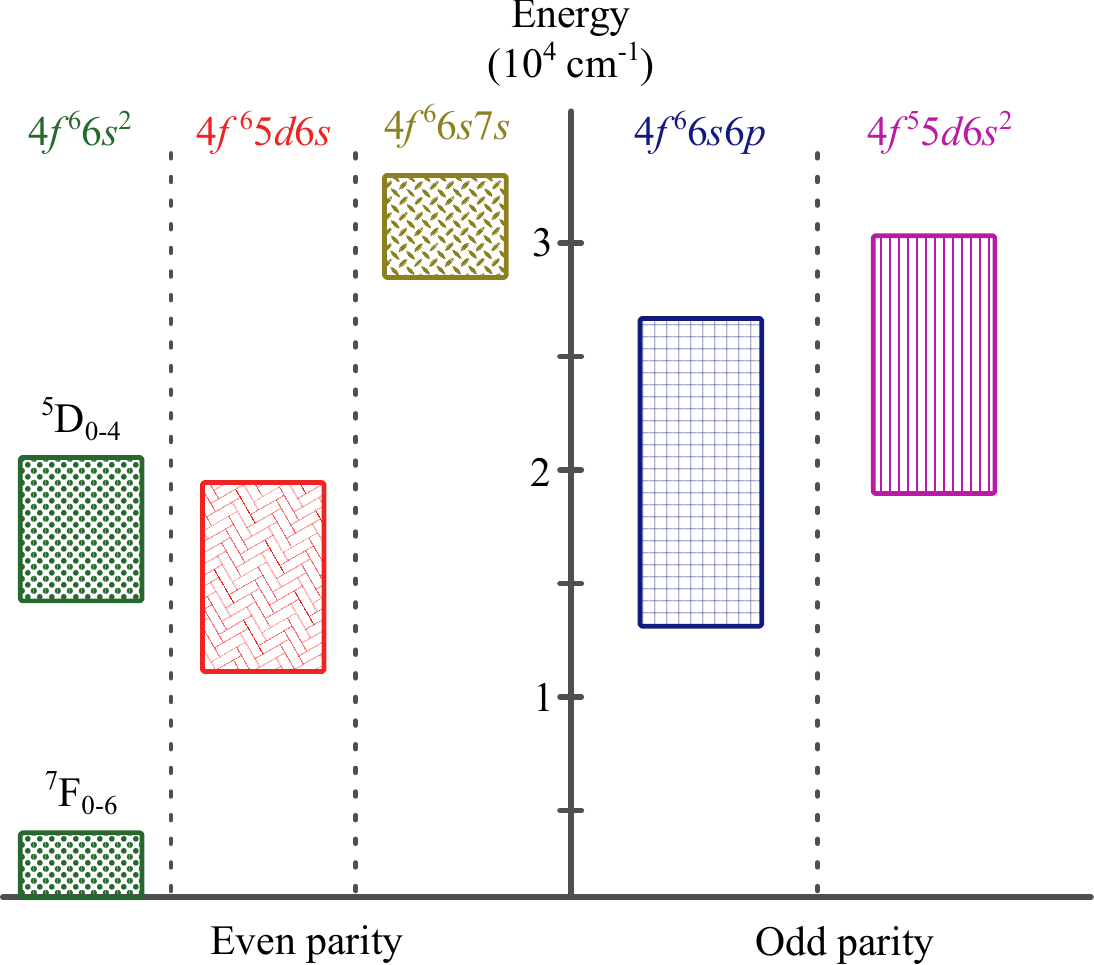}
	\caption{Known and relevant electronic configurations of samarium and their corresponding energy ranges. The ionization limit \cite{NIST_Dataset} is 45\,519.69\,cm$^{-1}$. The $^5$D levels indicated in the figure are designated in Ref.\,\cite{NIST_Dataset} as $^5$D3.}
	\label{Fig:levels_full}
\end{figure}
\begin{figure*}[t!]
	\centering
	\includegraphics[width=1\linewidth]{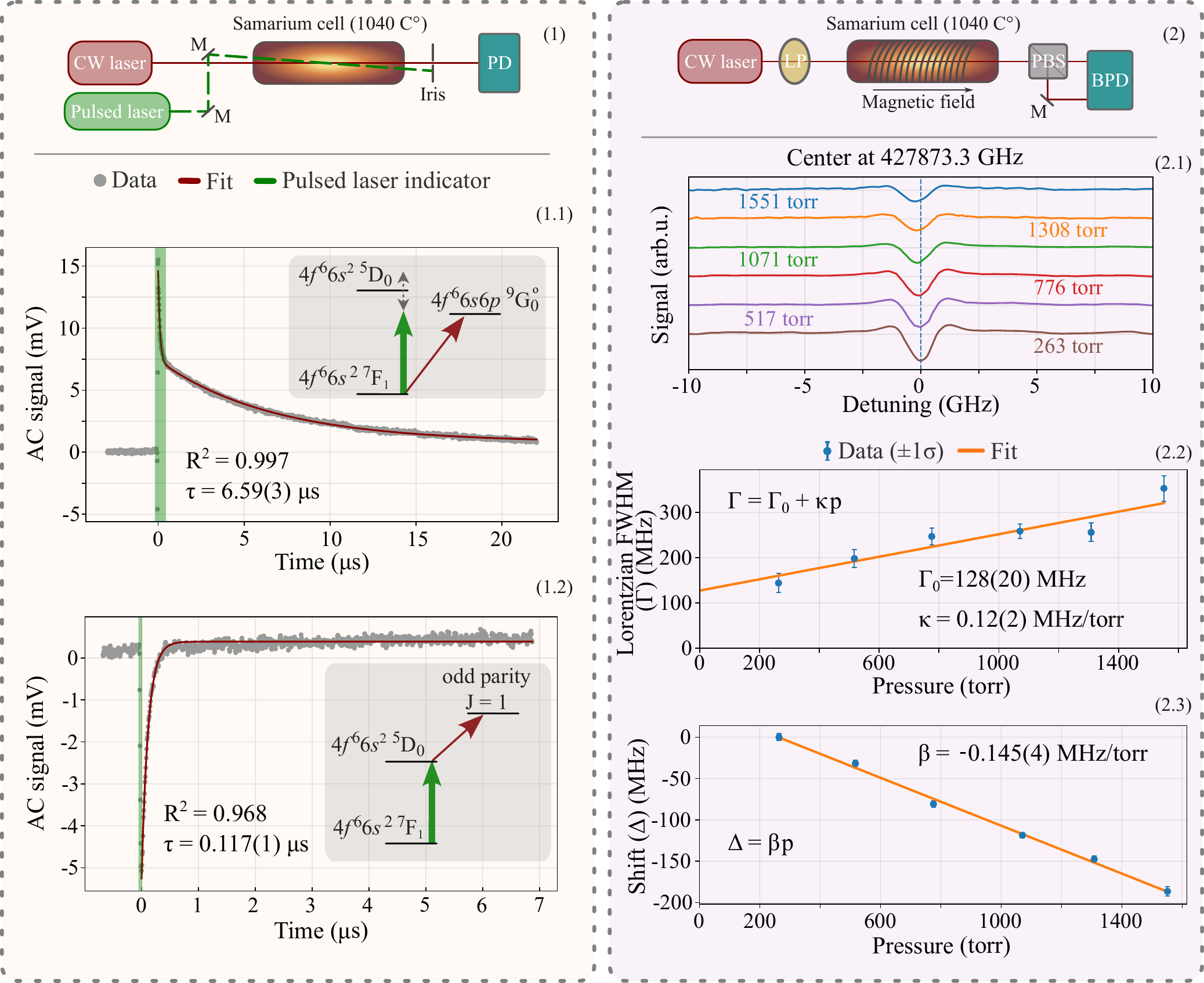}
	\caption{
    Spectroscopic level assignment and pressure-broadening measurement. (1) Setup for population-depletion and sequential-excitation spectroscopy. In (1.1), the positive transient signal reflects depletion of the $4f^66s^2\,{}^7F_1$ population; in (1.2), the negative transient signal reflects induced absorption from the populated candidate upper state. The zero level denotes the AC baseline of the CW-probe signal. The decay times are likely limited by atomic transit in (1.1) and by optical pumping/depletion in (1.2). (2) Setup for pressure-dependent Faraday-rotation spectroscopy. All pressures were measured with an uncertainty of $\pm15$~torr. M: mirror; PD: photodetector; LP: linear polarizer; PBS: polarizing beam splitter; BPD: balanced photodetector.
}
	\label{Fig:setups_results}
\end{figure*}

In the 1980s, it was suggested \cite{Dzuba1986} that the low-lying states $4f^{6}6s^{2}\,^{5}$D$_{J}~(J=0-4)$, not experimentally identified at that time, could be of interest for parity violation (PV) experiments due to a possible enhancement of the PV  effects. 
This interest is motivated by the possible existence of a nearby opposite-parity level (see Fig.\,\ref{Fig:levels_full}) and by the presence of a long isotopic chain that includes both deformed and spherical spin-zero nuclei, as well as isotopes with nonzero nuclear spin, enabling a broad range of PV studies, including measurements of nuclear spin-dependent (NSD) PV effects. 

Motivated by this proposal, a search for these states was undertaken, and three of them $^{5}$D$_{1}$, $^{5}$D$_{2}$, $^{5}$D$_{3}$ were found \cite{Barkov1988}.

However, experimental attempts to measure PV in the identified $4f^{6}6s^{2}\,^{7}$F$\rightarrow 4f^{6}6s^{2}\,^{5}$D magnetic-dipole (M1) transitions were unsuccessful \cite{Davies1989,Wolfenden1993}. This was later explained \cite{Rochester1999Sm_lifetimes} by noting that the expectations of PV enhancement had been based on erroneous identifications of the configuration and term assignments of some of the Sm I levels, which have since been corrected in \cite{NIST_Dataset}.

The energies of the $4f^{6}6s^{2}\,{}^{5}$D$_{0}$ and $^{5}$D$_{4}$ states in samarium were hitherto experimentally unknown. The identification of these levels could reinvigorate the search for atomic PV in Sm due to a particularly attractive experimental scheme suggested below. 

In particular, the $^{5}$D$_{0}$ state is a low-lying level~\cite{Beck|2010|Improved_prediction}. It has only a few possible electric-dipole (E1) decay channels, with small energy separations on the order of a few hundred $\mathrm{cm^{-1}}$. This suggests that this is likely a long-lived metastable level. In addition, transitions within the $f$ shell are well shielded from environmental perturbations.
Taken together with the existence of nearby opposite-parity levels, these properties make the $^{5}$D$_{0}$ state particularly promising for applications in atomic clocks, high-precision spectroscopy, isotope-shift measurements, and atomic parity violation (APV) experiments.

\textit{Experiment:} In this work, we identified the  ${}^{5}$D$_{0}$ level of samarium by starting with the candidate lines in the recorded dual-frequency-comb absorption spectra of an atomic vapor \cite{Aramyan2025} that did not match transitions between the known levels \cite{NIST_Dataset}.
We identified the $4f^{6}6s^{2}\,^{7}$F$_1\rightarrow 4f^{6}6s^{2}\,^{5}$D$_0$ M1 transition and its upper 
level at $14\,564.90(2)\,\mathrm{cm^{-1}}$ using double-resonance population-depletion and sequential excitation spectroscopy, see Fig.\,\ref{Fig:setups_results}(1), to filter transitions from the $^7$F$_1$ to upper states with even parity, and Faraday rotation spectroscopy to measure the pressure broadening and shift of the $\mathrm{{}^7F \rightarrow {}^5D}$ transition as, correspondingly, $\approx$ 0.12\,MHz/torr and $\approx$ -0.145\,MHz/torr. Details on the experimental methods can be found in Sec. \ref{subsec:Experiment}. 

\textit{Theory:} 
There are two efficient approaches to the calculations for complex atoms with many electrons in open shells: diagonalization of the large-size configuration interaction (CI) matrix~\cite{CheKozPor25} 
and combination of the CI method with perturbation theory (CI+PT), where the contribution of the states with high energy is treated using many-body perturbation theory methods \cite{DzubaFlKozlov1996,CIPT}. Results of both approaches for Sm PV agree with reasonable accuracy. Details of the calculations are presented in Sec.\,\ref{subsec:Theory}.

\textit{Parity violation:}
APV was the main original motivation for searching for the $4f^66s^2\,^5$D states \cite{Dzuba1986,Barkov1988}. The analysis of the spectrum was considerably revised in light of the results \cite{Rochester1999Sm_lifetimes}, which explained why the early attempts \cite{Warrington|1995|PNC_Bismuth_Samarium,Lucas|1998|PNC_opt_rotation,Wolfenden1993,Davies1989} to measure parity-violating optical rotation on the $4f^66s^2\,^7$F$ \rightarrow 4f^66s^2\,^5$D transitions found in \cite{Barkov1988,barkov1989_Sm_levels_OS} were unsuccessful.
The presence of nearby opposite-parity $J=0$ and $J=1$ states close to the
$4f^66s^2\,{}^5$D$_0$ level enables studies of both
nuclear-spin-independent (NSI) and NSD PV. The
$4f^66s^2\,{}^7$F$_1 \rightarrow 4f^66s^2\,{}^5$D$_0$ transition can contain
both NSI- and NSD-PV contributions, whereas the
$4f^66s^2\,{}^7$F$_0 \rightarrow 4f^66s^2\,{}^5$D$_0$ transition is free from
NSI-PV contributions to first order.
Although NSD PV can in principle be extracted in the presence of a typically two orders of magnitude larger NSI contribution, the absence of the dominant NSI-PV amplitude provides a cleaner probe and reduces the reliance on detailed modeling of the NSI background, for example, its lineshape. Moreover, despite its fundamental importance, NSD PV has so far been measured only once, in cesium \cite{WooBenCho97}. Therefore, the $4f^66s^2\,{}^7$F$_0 \rightarrow 4f^66s^2\,{}^5$D$_0$ transition amplitude is the central focus of this work.

The Hamiltonian describing the nuclear spin-dependent PV electron-nuclear interaction can be presented as:
\begin{eqnarray}
H_{\rm NSD} = \frac{G_F}{\sqrt{2}} \frac{\kappa}{I} {\bm \alpha} {\bm I} \rho({\bm r})\,,
\label{Eq:NSD}
\end{eqnarray}
where $G_F \approx 2.2 \times 10^{-14}$ a.u. is 
the Fermi constant of the weak interaction, $\kappa$ is the dimensionless coupling constant,
$\bm\alpha$
are Dirac matrices, $\bm I$ is the nuclear spin, and $\rho({\bf r})$ is the nuclear density distribution. There are three contributions to the interaction constant $\kappa$: the nuclear anapole moment \cite{FlaKhr80,FlaKhrSus84,FlaMur97}, Z-boson exchange between electron and nucleus \cite{FlaNovSusKhr1977}, and a combination of the NSD interaction mediated by Z-boson exchange (proportional to the nuclear weak charge) and hyperfine interaction \cite{FlaKhr1985}. The formulas and values for all three  contributions are presented in     
 Sec.\,\ref{subsec:NSD-constant}.

The nuclear anapole moment is produced by PV nuclear forces and is expected to give a dominant contribution to NSD PV in atoms. Thus, we have a possibility to investigate parity-violating nuclear forces in atomic experiments.

PV nuclear forces are usually described in terms of the meson exchange theory.
In the case of a valence proton (like in $^{133}$Cs), the PV nuclear potential is proportional to the sum of the $\rho$-meson and pion contributions, while in the case of a valence neutron (like in  $^{147}$Sm and $^{149}$Sm) this is the difference of $\rho$-meson and pion contributions. This makes the neutron case especially sensitive to the pion-nucleon PV interaction constant. The value of this constant was debated for many years, and it is still poorly known. Here, a Sm PV experiment can give an important result.    
 
Our calculations presented in detail in Sec.\,\ref{subsubsec:PV amplitude} give the following value of the reduced NSD PV transition amplitude:  
 \begin{equation}
\langle ^5\mathrm{D}_0\,I || d_{\rm NSD} || ^7\mathrm{F}_0\, I \rangle \approx 3 \times 10^{-13}\, i \kappa \, |e| a_0\,.
\label{Eq:E_PNC}
\end{equation}
This value is comparable to NSD contribution to PV amplitude in Cs \cite{WooBenCho97,FlaMur97}.
Note that the ratio of PV effects in $^{147}$Sm and $^{149}$Sm is insensitive to uncertainties in atomic calculations. This may be an important advantage compared to Cs, which has only one stable isotope.

\begin{figure*}[t]
	\centering
    \includegraphics[width=1\linewidth]{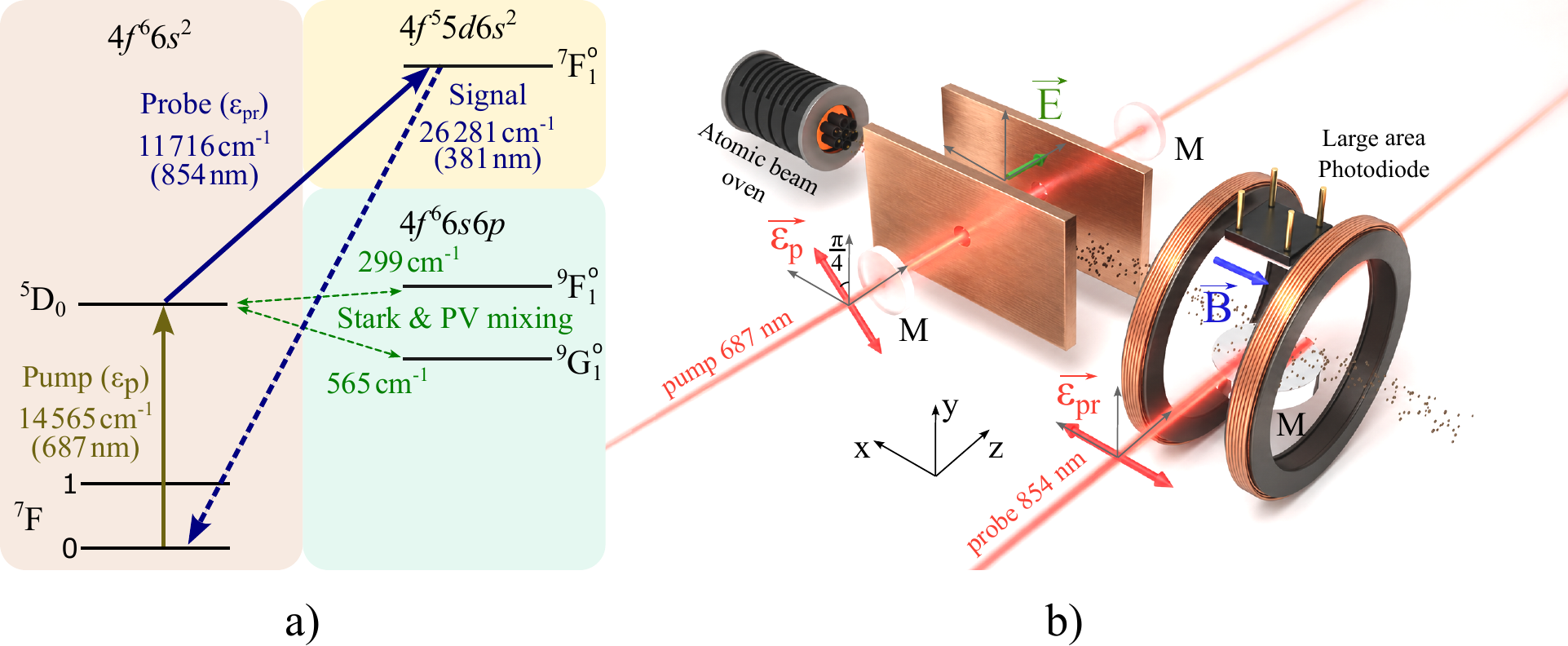}
	\caption{(a) Energy level diagram of samarium illustrating the atomic states relevant to the APV experiment, along with their energy separations. (b) Proposed experimental geometry: an atomic beam interacts with the pump ($\vec{\varepsilon}_p$) and probe ($\vec{\varepsilon}_{pr}$) fields in the presence of applied electric ($\vec{E}$) and magnetic ($\vec{B}$) fields; the 381\,nm fluorescence signal is detected with a photodiode. M: mirror; the mirrors adjacent to the electrodes form a power-buildup cavity to enhance the optical path length.}
	\label{Fig:Sm_PV}
\end{figure*}

We propose an experiment based on the Stark interference technique introduced by the Bouchiats \cite{Bouchiat|1975|PVpart2} and successfully employed by the JILA group \cite{WooBenCho97,Wood|1999|CJP_Cs_PNC} in the most precise APV experiment with Cs and by our group in Mainz \cite{Antypas|2018|APV_Yb,Antypas|2019|PRA_APV_Yb} for the measurement of APV in an isotopic chain of Yb.

We suggest searching for a parity-violating transition amplitude of the highly forbidden $4f^66s^2\,^7$F$_0 \rightarrow 4f^66s^2\,^5$D$_0$ transition. It becomes weakly allowed due to PV and Stark mixing of the  $4f^66s^2\,^5$D$_0$ state with the $4f^66s6p\,^9$F$_1$ odd-parity state 299\,cm$^{-1}$ above (Fig.\,\ref{Fig:Sm_PV}\,a). A particularly attractive feature of this transition is that there are no background M1 or E2 amplitudes of transitions between states with $J=0$. There are small M1 and E2 amplitudes that arise due to hyperfine interactions that mix the states (in particular  $4f^66s^2\,^5$D$_0$) with states of the same (even) parity and $J=1$ and 2 (see Sec.\,\ref{Sec:HFI_amplitudes}). These are additionally suppressed due to the smallness of the hyperfine constants in the $4f^66s^2$ states \cite{Childs|1972|HFS_Sm}. 
The ratio of the NSD-PV amplitude to the hyperfine-structure (HFS)-induced background amplitude in Sm is more than two orders of magnitude larger than the ratio of the NSD-PV amplitude to the background amplitude in Cs.  This should significantly reduce systematic effects related to the background amplitude.

A possible experimental scheme can be similar to those of the C.\,Wieman group experiment \cite{Wood|1999|CJP_Cs_PNC} with Cs  relying on the rotational invariant $\vec{\sigma}\cdot \vec{E}\times\vec{B}$ or the Berkeley/Mainz experiment \cite{Antypas|2019|PRA_APV_Yb} with Yb with the rotational invariant 
\begin{equation}
(\vec{\epsilon}\cdot \vec{B})(\vec{\epsilon}\cdot\vec{E}\times\vec{B})\,.
\end{equation}
Here $\vec{\sigma}$ and $\vec{\epsilon}$ describe circular and linear polarizations of the light exciting the respective forbidden transitions and $\vec{E}$ and $\vec{B}$ are (quasi)static electric and magnetic fields applied to the atoms. The form of these rotational invariants can be traced to their origin, related to the interference of the PV and Stark-induced transition amplitudes. The reason for the magnetic field appearing in these invariants is that the PV effect would average to zero if summed over all Zeeman components \cite{Khr91}. 

In the present case, because $J=0$ levels only shift due to nuclear magnetism, considerable Zeeman shifts may only be achieved during the probe stage of the experiments when the population of the  $4f^66s^2\,^5$D$_0$ is analyzed by laser excitation to a higher state detection of fluorescence, see a possible scheme depicted in Fig.\,\ref{Fig:Sm_PV}a.

There is also an alternative to using a magnetic field that corresponds to the rotational invariant
\begin{equation}
(\vec{\epsilon_p}\cdot \vec{\epsilon_{pr}})(\vec{\epsilon}_p\cdot\vec{E}\times\vec{\epsilon}_{pr})\,.
\end{equation}
Here $\vec{\epsilon}_p$ and $\vec{\epsilon}_{pr}$ are the light-polarization vectors for the pump and probe light, respectively. A corresponding experiment is sketched in Fig.\,\ref{Fig:Sm_PV}b.

In contrast to the transitions in other atomic systems where PV has been studied before, the PV amplitude of the Sm $4f^66s^2\,^7$F$_0 \rightarrow 4f^66s^2\,^5$D$_0$ transition arises exclusively due to the NSD PV effects, which are typically two orders of magnitude smaller than the (here absent) nuclear spin-independent amplitude. Indeed, weak interactions only mix states of the same total angular momentum $F$ and its projection $m_F$. In the absence of nuclear spin $I$ the mixing is between $J=0$ states and the E1 amplitude is forbidden. 

The measurement in the Sm system can be strongly motivated by the following:
\begin{itemize}
    \item Thus far, NSD PV has only been detected in a single $^{133}$Cs system and only in one experiment \cite{WooBenCho97}. 
    \item In Sm, there are two stable isotopes, $^{147}$Sm and $^{149}$Sm (the nuclear spin is $I=7/2$ for both), where measurements can be performed in the same experiment and compared.
Note that the ratio of the measured values of $\kappa$ in $^{147}$Sm and $^{149}$Sm is insensitive to errors in atomic calculations of the PV amplitude. This will provide a clean test of the theory of \textit{nuclear} PV.
    \item In both these isotopes, the unpaired nucleon is a neutron, as opposed to an unpaired proton in Cs. These systems are thus complementary. 
    \item Systematic effects in the Sm experiment are expected to be more tractable than in other systems due to the absence of the nuclear spin-independent PV background and the relative smallness of the background M1 and E2 transition amplitudes (at least two orders of magnitude smaller than in Cs) that are nonzero only due to hyperfine interaction-induced mixing. An additional handle is that, for example, the induced E2 transition amplitudes and the associated systematics should be proportional to the quadrupole moments of $^{147,149}$Sm, which are of opposite sign and different in magnitude by a factor of 3.6; see Sec.\,\ref{Sec:HFI_amplitudes}. 
\end{itemize}

\medskip
Beyond its application to the NSD PV experiment, the newly identified state opens further opportunities for precision measurements and searches for physics beyond the Standard Model, including atomic clocks, nonlinear isotope shifts, and searches for variation of the fine-structure constant $\alpha$, as discussed below. Possible laser-cooling schemes for atomic samarium are also considered.

\textit{Atomic clocks:} 
There is only one possible E1 decay channel from the newly found $^5$D$_0$ state, namely to the ${}^9$G$_1^\circ$ state, which contains approximately a $4\%$ admixture of the ${}^7$F$^\circ$ term \cite{NIST_Dataset}. 
The calculated rate of this decay is much lower 
than that of the M1 decay back to the $\mathrm{{}^7F_1}$ state. The calculation carried out within the [9spdfg] CI approximation yields a total lifetime of about 120\,ms while an independent CI+PT calculation gives about 200\,ms. We use the difference between the two values as an estimate of the theoretical uncertainty.
This corresponds to a natural linewidth order of 1\,Hz. 
With the M1-transition frequency corresponding to $14\,272.32\,\mathrm{cm^{-1}}$ 
($\nu \approx 4.28\times10^{14}\,\mathrm{Hz}$), the resulting quality factor is 
$\mathcal{Q} = \nu/\Delta\nu\sim 10^{14}$, typical for atomic lattice clocks \cite{Ludlow|2015|RevModPhys_Optical_clocks}.

The most accurate lattice clocks \cite{Aeppli|2024|Sr_clock_10_19,Hinkley|2013|Yb_10_18} utilize $J=0 \rightarrow J'=0$ forbidden transitions in Sr and Yb that become weakly allowed due to HFS-induced mixing. 
Such transitions are insensitive to first-order Zeeman shifts. However, the clock transitions $5s^2 \rightarrow 5s5p$ in Sr, and $6s^2 \rightarrow 6s6p$ in Yb involve excitation of outer-shell electrons, which are not shielded from external perturbations.
As a result, their performance is not limited by the natural linewidth of the transition but rather by environmental effects such as blackbody radiation (BBR), electric-field (Stark) shifts and collisional interactions.
In this context, the Yb $4f \rightarrow 5d$ transition has attracted significant interest in recent years \cite{Dzuba|2018|4f_to_5d_Yb,Ishiyama|2023|Inner-Shell_clock_transition_Yb,Ishiyama|2026|Improvement_Yb_innershell}, owing to its partial shielding and consequently relatively reduced sensitivity to external fields and its relevance to searches for physics beyond the Standard Model.

The Sm transitions from the ground multiplet to the ${}^{5}$D multiplet occur entirely within the $4f$ shell, whose electrons are shielded from external perturbations, resulting in reduced sensitivity to environmental perturbations. This shielding, together with the long lifetime of the ${}^{5}$D$_0$ state, makes this transition a promising candidate for optical-clock applications. A similar example is thulium, where the $4f^{13}6s^2\,{}^{2}$F$^\circ_{7/2} \rightarrow 4f^{13}6s^2\,{}^{2}$F$^\circ_{5/2}$ transition exhibits a low blackbody-radiation (BBR) shift \cite{Golovizin|2019|Tm_clock}. 
This shielding is also conceptually analogous to nuclear clocks, where the nuclear transition is screened from external perturbations by the electronic shell \cite{Beeks|2021|Th_clock}.
As an example, when samarium ions are embedded in a host solid matrix (e.g., SrF$_2$), the transition exhibits low differential sensitivity to crystal-field effects, and correspondingly low inhomogeneous broadening \cite{Verma2019}.
It is therefore promising for applications in portable solid-state atomic clocks. An exact analog of this Sm transition also occurs in Eu$^{3+}$ ions in solids, where the  $4f^{6}\,{}^{7}$F$_{0} \rightarrow 4f^{6}\,^5$D$_0$ transition has an exceptionally narrow linewidth allowing one to resolve the hyperfine structure \cite{Ahlefeldt|2016|Eu_first}, allowing for optical pumping of nuclear hyperpolarization \cite{Sushkov|2023|Eu+3}, and searches for physics beyond the Standard Model \cite{Nima|2026|Limit_Schiff_moment_Eu}.

\begin{figure}[h!]
	\centering
	\includegraphics[width=1\linewidth]{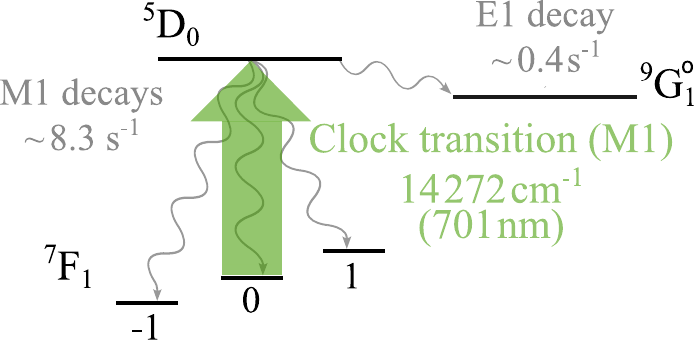}
    \caption{Candidate clock transition at $14\,272\,\mathrm{cm}^{-1}$ ($701\,\mathrm{nm}$), together with the competing M1 and E1 decay channels from the $^{5}$D$_{0}$ state to the $^{7}$F$_{1}$ and $^{9}$G$_{1}^{\circ}$ states, respectively. Indicated are the decay rates between the corresponding levels summed over the lower-state Zeeman sublevels, obtained from CI calculations.}
	\label{Fig:Sm_Clock}
\end{figure}

In contrast to optical lattice clocks based on transitions between opposite-parity states, where the $J=0 \rightarrow J'=0$ transition acquires an E1 amplitude through HFS mixing, the same-parity transition considered here can gain strength via M1 or E2 admixture from nearby same-parity states. The nearby ${}^{7}$P$_2$ state, separated by only $14.4\,\mathrm{cm}^{-1}$, is expected to provide an enhanced contribution to the hyperfine-induced E2 amplitude due to the small energy denominator. Other nearby states, with energy separations on the order of hundreds of $\mathrm{cm}^{-1}$, may also contribute, although their larger energy separations suppress the corresponding mixing amplitudes. The resulting hyperfine-induced ${}^{7}$F$_0 \rightarrow {}^{5}$D$_0$ E2 transition is present only in isotopes with nonzero nuclear spin, namely $\mathrm{^{147}Sm}$ and $\mathrm{^{149}Sm}$, which have natural abundances of $15\%$ and $13.8\%$, respectively.

Alternatively, the directly M1-allowed $\mathrm{{}^7F_1 \rightarrow {}^5D_0}$ transition can be used for clock interrogation by applying an external magnetic field to resolve the Zeeman components, see Fig.\,\ref{Fig:Sm_Clock}. This enables the selection of the $m_J=0 \rightarrow m'_J=0$ component, which is insensitive to the first-order Zeeman effect. In this case, direct M1 interrogation avoids the need for hyperfine-induced E2 excitation and provides a stronger transition amplitude. Moreover, it allows the use of the more abundant even isotopes, potentially enabling a higher signal-to-noise ratio.
Repumping among the Zeeman sublevels can be achieved via excitation on the 
$\mathrm{{}^7F_1 \rightarrow {}^7F_0^\circ}$ transition at 459.8\,nm, 
with the odd-parity upper level at $22\,041.02\,\mathrm{cm^{-1}}$. This transition is among the strongest in the samarium spectrum, with an experimentally measured transition probability of $96(5)\times10^6~\mathrm{s^{-1}}$~\cite{Lawler2013}. 
The $\mathrm{{}^7F_1 \rightarrow {}^7F_0^\circ}$ transition may also be suitable for laser cooling, which is required for lattice-clock applications involving trapped Sm atoms. We have calculated the decay rates from the ${}^7F_0^\circ$ state and find that the dominant decay channel is to the ${}^7F_1$ ground state. In addition, there are seven weaker decay channels to other states, which are discussed in Sec.\,\ref{Sec: Laser cooling}.

Cooling is essential for a number of experiments, such as isotope-shift measurements. 
These have recently attracted significant interest, particularly in studies of nonlinear effects in the King plot for isotope shifts (IS) as probes of new physics \cite{Berengut2018,Door|2025|Ytterbium_IS_New_Bosons,Berengut|2025|IS}. 
However, such searches may be hindered by ``old-physics'' nonlinearities, which, in fact, were first observed and studied in Sm\,I \cite{Griffith1981} and explained by second-order IS effects. 
The $^7$F$_1 \rightarrow {}^5$D$_0$ transition is particularly interesting for this type of research, as the field shift is expected to be negligible, so that the isotope shift is expected to be dominated by the (anomalous) mass shift. Such mass shifts are a probe of electron correlations \cite{King|1984|IS_book}.

\textit{Variation of fundamental constants:}
Since the late 1990s, research attention has been drawn to testing the temporal stability of fundamental constants such as the fine structure constant $\alpha$, see, for example, reviews \cite{Safronova2018,Flambaum|2026|Alpha}. Both slow drift on astronomical time scales and fast oscillations \cite{Antypas|2020|Fast_oscil} have been considered, which may arise due to the presence of an ultralight bosonic dark matter field \cite{Stadnik|2015|Dark_matter_alpha}. One possible manifestation of the variation of $\alpha$ with respect to its reference value $\alpha_0$ is a change in the energy of an atomic state $\mathcal{E}_i$:
\begin{equation}\label{eq:Alpha_variation}
\Delta \mathcal{E}_i = q_i\cdot\left [ \left(\frac{\alpha}{\alpha_0}\right )^2-1 \right ]\,,
\end{equation}
where $q_i$ is the sensitivity coefficient for the corresponding state. The calculated sensitivity coefficients for several states of Sm are listed in Table.\,\ref{tab:q_sensitivity}.

\begin{table}[h!]
    \centering
    \caption{Calculated coefficients for the sensitivity of the energies of the listed states to variations of the fine-structure constant $\alpha$. The uncertainty of the calculation is $\approx 10\%$.}
    \begin{tabular*}{\linewidth}{@{\extracolsep{\fill}} c c c c}
        \toprule\toprule
        Config. & Term & Energy (cm$^{-1}$) & $q_i$ (cm$^{-1}$) \\
        \midrule
        $4f^66s^2$ & $^7$F$_0$ & 0 & 0 \\
        $4f^66s^2$ & $^7$F$_1$ & 292.58 & 321 \\
        $4f^66s5d$ & $^9$H$_1$ & 10\,801.10 & 5111 \\
        $4f^66s5d$ & $^7$P$_2$ & 14\,550.50 & 6091 \\
        $4f^66s^2$ & $^5$D$_0$ & 14\,564.90 & -3529 \\
        \bottomrule\bottomrule
    \end{tabular*}     
    \label{tab:q_sensitivity}
\end{table}
A possible experiment taking advantage of the newly found $^5$D$_0$ state could be a precise frequency comparison of the $4f^66s^2\,^7$F$_1 \rightarrow 4f^66s^2\,^5$D$_0$ M1 transition and the $4f^66s^2\,^7$F$_1 \rightarrow 4f^65d6s\,^7$P$_2$ E2 transition. 
The lifetime of the $4f^6 6s^2\,\,{}^5$D$_0$ state is calculated to be $120-200\,\mathrm{ms}$, while
the lifetime of the $5d6s\,\,{}^7$P$_2$ state is
on the order of 1\,s. Thus, both states are long-lived, and the intrinsic linewidths of both transitions are narrow.

The energy separation between these two upper states (and thus the transition energies) is $\Delta \mathcal{E} = 14.4\,\mathrm{cm^{-1}}$.
For the detection of slow variations, a particularly attractive possibility is the realization of a dual Sm atomic clock based on two clock transitions sharing the same lower state. Comparing the outputs of the two clocks yields a differential sensitivity
\begin{equation}
    q = q(^7P_2)-q(^5D_0)=9620\,\mathrm{cm}^{-1}.
\end{equation}
Such a scheme benefits from common experimental infrastructure, reduced sensitivity to shifts of the shared lower state, and the involvement of shielded electronic shells, making the transitions comparatively insensitive to external perturbations.
Another possibility is direct spectroscopy of the $0.43$\,THz transition between the $^5$D$_0$ and $^7$P$_2$ states. Due to the small transition frequency, this transition exhibits a large enhancement factor
\begin{equation}
    \mathcal{K} = \frac{2q}{\Delta\mathcal{E}}=-1336\,,
\end{equation}
which characterizes the sensitivity of the transition frequency $\omega$ to variation of the fine-structure constant $\alpha$, $\delta \omega/\omega=\mathcal{K} \delta \alpha/\alpha$. The low transition frequency would also result in a reduced Doppler width.
A possible scheme to exploit this enhancement is to measure the transition frequency via coherent Raman spectroscopy using, for example, the allowed $4f^66s^2 \rightarrow 4f^66s6p \rightarrow 4f^65d6s$ transitions.

This system can also be used for the search for fast variations of fundamental constants produced by a dark matter field \cite{Antypas|2020|Fast_oscil} along the lines of what was done in \cite{Tretiak2022_RFDMPRL} with a Cs transition. For example, using the $4f^66s^2\,^7$F$_1 \rightarrow 4f^66s^2\,^5$D$_0$ M1 transition,  a similar experiment will be more sensitive by a factor of $\approx$2.4 due to the greater sensitivity factor $q$, assuming otherwise identical experimental conditions.

\textit{Conclusion:}
We report the first experimental identification of the previously unobserved $4f^6 6s^2 \, ^5$D$_0$ level in neutral samarium at $14\,564.90(2)\,\mathrm{cm}^{-1}$. 
The observed small pressure broadening (0.12(2)\,MHz/torr) and shift (-0.145(4)\,MHz/torr) of a transition from the ground term to this level indicate that this transition indeed occurs within the $4f$ shell and is effectively shielded by outer electrons, resulting in reduced sensitivity to external perturbations.

Our theoretical analysis shows that this state has a single allowed E1 decay channel with a transition rate 
much smaller than the competing M1 decay back to the ground-term $^7$F$_1$ level, yielding a lifetime in the range $120$--$200\,\mathrm{ms}$.
The corresponding quality factor ($10^{14}$) of the fully shielded $4f \rightarrow 4f$ transition is comparable to those of state-of-the-art optical clock transitions.

Furthermore, the calculated reduced NSD-PV amplitude for the $4f^66s^2\,^7$F$_0 \rightarrow 4f^66s^2\,^5$D$_0$ transition (Eq.~\ref{Eq:E_PNC}) is comparable to that of Cs, the only system in which NSD PV has so far been observed and only once. Unlike in Cs, the Sm transition is free from NSI-PV contributions and has substantially suppressed parity-conserving backgrounds: the HFS-induced amplitudes are more than two orders of magnitude smaller than the corresponding background in Cs. The dependence of the dominating HFS-induced E2 amplitude on the nuclear quadrupole moments of $^{147,149}$Sm provides an additional handle for controlling systematics. Together with the proposed rotational invariants, these features make Sm a particularly clean system for future NSD-PV measurements.

The sensitivity coefficient of the $^5$D$_0$ state to variations of the fine-structure constant $\alpha$ is calculated to be $-3529\,\mathrm{cm}^{-1}$. Together with a nearby metastable same-parity level with a coefficient of $6091\,\mathrm{cm}^{-1}$, this provides a large differential sensitivity, making samarium a promising candidate for self-referenced searches for temporal variation of $\alpha$.

A possible laser-cooling transition has also been identified. The estimated leakage into other decay channels is only about $0.1\%$ of the total decay rate, suggesting that laser cooling of Sm may be achievable with additional repumping. Such a cooling scheme would enable trapped-samarium studies, including precise lifetime measurements, isotope-shift studies on transitions dominated by the anomalous mass shift, and the development of samarium-based optical frequency standards.

Overall, these results establish neutral samarium as a versatile platform for precision spectroscopy in complex open-$4f$ systems, with potential for applications in optical clocks, tests of fundamental symmetries, and searches for physics beyond the Standard Model.

{\def\MakeTextUppercase#1{#1}\section*{METHODS}}
\label{sec:methods}
\subsection{Experiment}
\label{subsec:Experiment}
In Ref.~\cite{Barkov1988}, the identification of the $4f^{6}6s^{2}\,{}^{5}$D$_{1,2,3}$ levels relied on searching for weak, previously unobserved transitions with small isotopic and hyperfine splitting and pressure broadening. 

The $\mathrm{{}^7F \rightarrow {}^5D}$ transitions are M1 transitions enabled by LS-coupling breakdown, with the ${}^5$D levels containing $\sim 20\%$ ${}^7$F admixture. Since M1 transition rates are orders of magnitude smaller than allowed E1 rates, these transitions are expected to be relatively weak.

Pressure broadening serves as the primary identifier: transitions occurring within the inner $4f$ shell are shielded from collisions by outer 5$s$, 5$p$, and 6$s$ electrons, resulting in collisional broadening at least an order of magnitude smaller than those of typical E1 transitions \cite{Vedenin|1987|Pressure_broadening}.

Here, we followed a similar approach, implementing several techniques guided by the theoretical prediction to narrow the search range.

The energy of the $^5$D$_0$ state was predicted~\cite{Beck|2010|Improved_prediction} to be approximately 
$14\,302\,\mathrm{cm^{-1}}$, implying that the
$\mathrm{{}^{7}F_{1}\!\rightarrow{}^{5}D_{0}}$ transition should occur near $14\,009\,\mathrm{cm^{-1}}$. However, even a few percent uncertainty in the prediction corresponds to frequency offsets of several hundred to about one thousand $\mathrm{cm^{-1}}$, i.e., a search range spanning several tens of terahertz.
This range was significantly reduced using the spectrum from our previous dual frequency-comb work~\cite{Aramyan2025}, which covers an interval between approximately $+5\%$ and $-2\%$ from the predicted frequency. This transforms the problem from a blind search into a targeted analysis of unidentified spectral features within the recorded data. 
The sensitivity of the system to such transitions was verified using the known $\mathrm{{}^{7}F_{2}\!\rightarrow{}^{5}D_{1}}$ transition, whose magnetic-dipole matrix element is $\langle f | \mathrm{M1} | i \rangle \approx -0.5\,\mu_B$, compared to $\approx -0.3\,\mu_B$ for the transition of interest \cite{Kra89}.

\begin{table*}[t]
\begingroup

\centering
\caption{Experimental evidence supporting the assignment of the observed level as
$4f^{6}6s^{2}\,{}^{5}$D$_{0}$.}
\label{tab:level-assignment}

\renewcommand{\arraystretch}{1.2}

\begin{tabular}{p{3cm} p{0.5cm} p{12cm}}
\hline
\textbf{Experiment} & & \textbf{Result} \\
\hline

\makecell[lt]{Dual-comb \\ spectroscopy} & &
Weak unidentified transitions are found within the predicted energy range of the
$4f^{6}6s^{2}\,{}^{5}$D$_{0}$ level.
\\

\makecell[lt]{Population-depletion \\ spectroscopy }& &
Selectively identifies transitions from
$4f^{6}6s^{2}\,{}^{7}\mathrm{F}_{1}$, excluding transitions from all other states, including thermally populated excited states.
\\

\makecell[lt]{Sequential-excitation \\ spectroscopy} & & 
Confirms the upper-state energy and even parity, excludes a two-photon transition, and narrows the possible assignments to
$4f^{6}5d6s\,{}^{7}$D$_{1}$ and
$4f^{6}6s^{2}\,{}^{5}$D$_{0}$.
\\

\makecell[lt]{Pressure-dependent \\ Faraday-rotation \\ spectroscopy }& &
The small pressure broadening and shift identify the transition as a shielded
$\mathrm{4f}\rightarrow\mathrm{4f}$ transition, excluding the
$\mathrm{4f}^{6}5d6s\,{}^{7}\mathrm{D}_{1}$ assignment and uniquely identifying the observed level as
$\mathrm{4f}^{6}6s^{2}\,{}^{5}\mathrm{D}_{0}$.
\\

\hline
\end{tabular}

\endgroup
\end{table*}

To identify the $^5$D$_0$ energy level, a sufficiently high samarium vapor density is required; therefore, the cell was heated to $\approx1040\,^\circ\mathrm{C}$. This heating results in a weak thermal population of excited states and many additional Boltzmann-factor-limited transition lines originating from these states in the spectrum. Since the Sm spectrum is not fully characterized, many of these lines are absent from existing datasets \cite{Ferrara|2023|Spectral_lines,Martin|1978|Energy_levels,NIST_Dataset,Meggers|1975|Spectral_lines} and may therefore be misidentified as candidates or obscure the transition of interest due to their proximity in transition energy.

Double-resonance population-depletion spectroscopy, see Fig.\,\ref{Fig:setups_results}(1), was implemented to selectively probe transitions originating from the $^7$F$_1$ state. A nanosecond pulsed laser (Cobra Stretch dye laser pumped with a Boston 500 Nd:YAG laser, Sirah Lasertechnik), with a pulse energy of $\approx$ 2\,mJ and a beam diameter of $\approx$ 2\,mm, was used to excite candidate transitions. The population of the $\mathrm{^{7}F_{1}}$ state was monitored using a continuous-wave (CW) laser (SolsTiS Ti:sapphire laser, M2) with a power of $\approx 5\,\mathrm{mW}$, tuned to the $4f^{6}6s^{2}\,{}^{7}$F$_{1} \rightarrow 4f^{6}6s6p{}\,^{9}$G$^{\mathrm{o}}_{0}$ (with 4$\%$ admixture of $\mathrm{^7F_0^\circ}$) transition. 
When the pulsed laser is tuned into resonance with a transition originating from the $\mathrm{{}^7F_1}$ state, it depletes the population of this level, thereby reducing the probe-light absorption, see Fig.\,\ref{Fig:setups_results}(1.1). The plotted signal corresponds to the AC component and is therefore zero off resonance.
This confirms that the transition originates from the $^{7}$F$_{1}$ state. However, the identity of the upper state is initially unknown. Therefore, the observed resonance could correspond to a two-photon transition or to excitation of another unidentified level, such as the $4f^65d6s\,^7$D$_1$ state.

To determine the parity and confirm the energy of the upper state, a sequential excitation scheme was employed using the same setup, see Fig.\,\ref{Fig:setups_results}(1), with the modification that the CW laser was tuned into resonance between the detected state and a known odd-parity state. The electronic configuration of the final state is unknown; however, the state is known to have odd parity and total angular momentum $J = 1$. A transition from this state to the $^{5}$D$_{1}$ level has been observed \cite{Ferrara|2023|Spectral_lines}.

Therefore, if the atoms are excited from the $^7$F$_1$ state to an even-parity state, which could be the $^{5}$D$_{0}$ state, a transient absorption signal should be observed with a properly tuned CW laser. This behavior is indeed observed, as shown in Fig.\,\ref{Fig:setups_results}(1.2).
This confirms the parity and the energy of the upper level, excluding the possibility of a two-photon transition. Consequently, only two assignments remain possible: $4f^65d6s\,^7$D$_1$ or $4f^66s^2~^5$D$_0$, as these are the only unidentified low-energy levels that can be excited from the $^7$F$_1$ state.

To confirm that this state is indeed $^5$D$_0$, it remained to show that the excitation occurs within the 4$f$ shell. To this end, we measured pressure broadening of this transition line as had been done in Ref.\,\cite{Barkov1988}.

The measurements were based on the Macaluso-Corbino effect, i.e., the Faraday rotation of the polarization plane in the vicinity of an atomic resonance~\cite{Budker2002}, see Fig.\,\ref{Fig:setups_results}\,(2). The CW laser beam (from the same Ti:sapphire laser), with a power of $\approx$ 3\,mW (increased to $10\,\mathrm{mW}$ at $1308$ and $1551\,\mathrm{torr}$ to compensate for convection-induced optical losses in the cell), propagates through a linear polarizer and an analyzer with their transmission axes oriented at $\approx 45^\circ$, with the samarium cell (see Appendix of Ref.\,\cite{Aramyan2025} for details) placed between them.
A magnetic field of $\approx 260\,\mathrm{G}$ was applied along the laser propagation direction.
The induced Zeeman shifts lead to different resonant frequencies for two circular polarizations $\sigma_{+}$ and $\sigma_{-}$ of input light, as well as different refractive indices $n_{+}$, $n_{-}$, resulting in Faraday rotation. 
The pressure broadening is determined by varying the pressure of the buffer gas (argon in our case) in the cell, 
Fig.\,\ref{Fig:setups_results}\,(2.1).
At each pressure, the resonance frequency and Lorentzian linewidth were extracted by fitting the measured Faraday-rotation spectra with a multicomponent model accounting for the unresolved even-isotope structure. The relative isotope positions, natural-abundance weights, Doppler width, and Zeeman splitting were fixed throughout the analysis, while the common transition frequency, Lorentzian linewidth, signal amplitude, and linear background were fitted. Because the isotope structure is not spectrally resolved, the individual component positions are not interpreted as measured isotope shifts, but rather provide a phenomenological description of the observed asymmetric line profile. The extracted transition frequencies and Lorentzian linewidths were subsequently analyzed as a function of pressure using weighted linear regression.

The transition at $14\,272.32\,(2)$\,cm$^{-1}$ exhibits relatively small pressure broadening of approximately $0.12(2)\,\mathrm{MHz/torr}$ and a pressure shift of approximately $-0.145(4)\,\mathrm{MHz/torr}$. These values are at least an order of magnitude smaller than those typically observed for transitions involving outer-shell electrons.

These results identify the observed transition as the $\mathrm{{}^7F_1 \rightarrow {}^5D_0}$ transition. The corresponding $^5$D$_0$ level is therefore determined to have an energy of $14\,564.90(2)\,\mathrm{cm^{-1}}$. The transition frequency was measured using a HighFinesse \AA ngstrom WS/6 wavemeter calibrated against a neon reference lamp, with a specified absolute accuracy of $600\,\mathrm{MHz}$ ($0.02\,\mathrm{cm^{-1}}$). The lower-state energy was taken from Ref.\,\cite{NIST_Dataset}, and an uncertainty of $0.01\,\mathrm{cm^{-1}}$ was assigned based on the numerical precision of the tabulated value. The quoted uncertainty of the level energy was obtained by combining the wavemeter calibration uncertainty and the uncertainty of the lower-state energy in quadrature. The statistical uncertainty of the spectral fits is at the MHz level and is therefore negligible on the scale of the quoted level energy.
Hence, taking into account the evidence summarized in Table~\ref{tab:level-assignment}, the energy of the $^5$D$_0$ level is determined to be $14\,564.90(2)\,\mathrm{cm}^{-1}$.

\subsection{Atomic theory}
\label{subsec:Theory}
\subsubsection{8-electron configuration interaction}
\label{CI}
We start from the solution of the Dirac-Fock equations. The Breit interaction is included at the self-consistent-field stage, while quantum electrodynamic (QED) corrections are neglected at this level of approximation. The even-parity states of interest belong to the configuration $4f^6 6s^2$, while odd-parity low-lying levels belong to two different configurations, $4f^6 6s6p$ and $4f^5 6s^2 5d$.

We carry out the initial self-consistency procedure for the [$1s^2$, ..., $4f^6 6s^2$]
configuration. Then, all electrons were frozen, and one electron from the $6s$ shell was moved to the $6p$ shell. Thus, the $6p$ orbital was constructed for the $4f^6 6s6p$ configuration. Subsequently, the orbitals are again frozen, and a two-step rearrangement is carried out: one electron is transferred from the $4f$ shell to the $6s$ shell, and another electron is promoted from the $6p$ shell to the $5d$ shell. Thus, the $5d$ orbitals were constructed for the $4f^5 6s^2 5d$ configuration.

The basis set employed in the CI calculations includes virtual orbitals up to $9s$, $9p$, $9d$, $9f$, and $9g$. These virtual orbitals are generated using the recursive procedure described in Refs.~\cite{Bogdanovich|1991|usage_transformed,KozPorFla|1996|nuclear_anapole_moment}, which ensures orthogonality to the occupied orbitals and proper asymptotic behavior. In practice, the size of the basis set is limited because the dimension of the CI space increases rapidly as higher-lying orbitals are included.

The CI configuration space is constructed by allowing single and double excitations from a set of reference configurations. For the even-parity states, excitations are generated from the $4f^6 6s^2$ and $4f^6 6s5d$ configurations, while for the odd-parity states, they are generated from the $4f^6 6s6p$ and $4f^5 6s^2 5d$
configurations. All excitations are restricted to the chosen virtual orbital space, ensuring a systematic and controlled expansion.

To investigate the convergence of the CI expansion, we perform calculations of the low-lying energy levels for four progressively enlarged configuration spaces:
(i) including only the two leading nonrelativistic configurations $4f^6 6s^2$ and $4f^6 6s5d$ for even-, and $4f^6 6s6p$ and $4f^5 6s^2 5d$ for odd-parity states;
(ii) including single and double excitations to the  $7s$, $6p$, $6d$, $6f$, and
$6g$ shells (we designate it [$7s6pdf\!g$]);
(iii) extending the excitation space to $[9spdf]$; and
(iv) including the $[9spdf\!g]$ virtual set.

In the largest calculation, case (iv), the CI space contains approximately 52 million Slater determinants for the even-parity states. This leads to a substantial computational cost and requires significant computational resources for the diagonalization of the Hamiltonian matrix. 

\subsubsection{Energy levels}
The low-lying energy levels were calculated using the four CI spaces described above. The results are presented in~\tref{tab_E}.

\begin{table}
\caption{The energy levels of the low-lying excited states counted from the ground state (in cm$^{-1}$).
The columns labeled ``two-conf.'', [$7s6pdf\!g$], [$9spdf$], and [$9spdf\!g$] give results obtained using different sets of the configurations
described in the text. The experimental energy levels~\cite{NIST_Dataset} are presented in the column labeled ``Experim.''. 
}
\label{tab_E}
\begin{ruledtabular}
\begin{tabular}{lccccccc}
 Config.      &     Term     & two-conf. & [$7s6pdf\!g$] & [$9spdf$] & [$9spdf\!g$] & Experim. \\
\hline \\ [-0.7pc]
$4f^6 6s^2$   & $^7$F$_0$    &    0     &      0      &     0     &       0    &    0   \\
$4f^6 6s^2$   & $^7$F$_1$    &          &             &           &      267   &   293  \\
$4f^6 6s^2$   & $^7$F$_2$    &          &             &           &      752   &   812  \\
$4f^6 6s^2$   & $^7$F$_3$    &          &             &           &     1397   &  1490  \\
$4f^6 6s^2$   & $^7\!F_4$    &          &             &           &     2155   &  2273  \\
$4f^6 6s^2$   & $^7$F$_5$    &          &             &           &     2989   &  3125  \\[0.2cm]
$4f^6 6s5d$   & $^9$G$_0$    &  19\,871   &    14\,995    &   14\,899   &    14\,982   & 13\,551 \\
$4f^6 6s^2$   & $^5$D$_0$    &  18\,443   &    16\,546    &   16\,379   &    15\,848   & 14\,565 \\
$4f^6 6s5d$   & $^7$F$_0$    &  23\,261   &    17\,554    &   17\,540   &    17\,607   & 15\,794 \\[0.3cm]
$4f^6 6s6p$   & $^9$G$_1^o$  &   7443   &    11\,117    &   12\,007   &    11\,789   & 14000 \\
$4f^6 6s6p$   & $^9$F$_1^o$  &   8386   &    12\,067    &   12\,996   &    12\,751   & 14\,864 \\
$4f^6 6s6p$   & $^7$G$_1^o$  &   9224   &    12\,814    &   13\,726   &    13\,485   & 15\,651 \\
$4f^6 6s6p$   & $^5$D$_1^o$  &   9706   &    13\,282    &   14\,204   &    13\,961   & 16\,112 \\
$4f^6 6s6p$   & $^7$D$_1^o$  &  10\,061   &    13\,718    &   14\,628   &    14\,386   & 16\,691 \\
\end{tabular}
\end{ruledtabular}
\end{table}

We were able to reproduce the low-lying even-parity states with $J=0$, belonging to the $4f^6 6s^2$ and $4f^6 6s5d$ configurations, reasonably well. However, the excited state $4f^6 6s^2\,\, ^5$D$_0$  calculated for [$9spdf\!g$] lies approximately 1300 cm$^{-1}$ higher than the experimental value.

As for the odd-parity states, we focus on those with $J=1$, since only such intermediate states contribute to the parity-violating amplitude. By comparing the experimental and calculated $g$-factors, we were able to identify the five lowest-lying odd-parity states with $J=1$, as listed in \tref{tab_E}. We note that the calculated excitation energies of these odd-parity states tend to be systematically lower than the experimental values.

We were not able to reproduce the relative positioning of the even- and odd-parity states accurately. This discrepancy may be attributed to two main factors: (i) the incomplete saturation of the CI space, and (ii) the limited treatment of core–valence correlations. The former arises from the rapid growth in the number of determinants when excitations to higher-lying shells are included. The latter is likely more significant, as the presence of eight valence electrons makes it impractical to apply methods that combine CI with many-body perturbation theory. Indeed, as seen from \tref{tab_E}, extending the CI space does not lead to a noticeable improvement in the energies of the low-lying states.

\subsubsection{M1 transition amplitudes}
To assess the quality of the wave functions for the $4f^6 6s^2\, ^7$F$_0$ and $4f^6 6s^2\, ^5$D$_0$ states, we calculated M$1$ transition amplitudes between levels of the $4f^6 6s^2$ configuration involving these states. The results of the calculation of the reduced matrix elements $|\langle J' ||\mu|| J \rangle|$ and the oscillator strengths $f(J \rightarrow J')$, given by the formula (in a.u.)
\begin{equation}
 f(J \rightarrow J') = -\frac{2\, \omega_{JJ'}}{3(2J+1)} |\langle J' ||\mu|| J \rangle|^2\,,
\end{equation}
where $\omega_{JJ'} \equiv E_J - E_{J'}$ is the transition frequency, are presented in \tref{tab_M1}.
\begin{table}
\caption{The absolute values of the reduced MEs $|\langle J' ||\mu|| J \rangle|$
(in Bohr magnetons $\mu_0$)
and the oscillator strengths $f(J \rightarrow J')$ (in a.u.) are presented for the transitions between the states belonging to the $4f^6 6s^2$ configuration.}
\label{tab_M1}
\begin{ruledtabular}
\begin{tabular}{lcccc}
& \multicolumn{2}{c}{$|\langle J' ||\mu|| J \rangle|$} & \multicolumn{2}{c}{$10^7 \times f(J \rightarrow J')$} \\
                                     & This work & \cite{Kra89} & This work & \cite{Beck|2010|Improved_prediction} \\
\hline \\ [-0.7pc]
$^7$F$_0 \rightarrow \,^7$F$_1 $     &     3.4   &              &   1.38    & 1.39  \\
$^7$F$_1 \rightarrow \,^7$F$_2 $     &     4.7   &              &   1.54    & 1.55  \\
$^7$F$_2 \rightarrow \,^7$F$_3 $     &     5.4   &              &   1.61    & 1.62  \\
$^7$F$_3 \rightarrow \,^7$F$_4 $     &     5.7   &              &   1.47    & 1.48  \\[0.2cm]
$^7$F$_0 \rightarrow \,^5$D$_1 $     &     0.13  &    0.160     &   11.0    & 11.6  \\
$^7$F$_1 \rightarrow \,^5$D$_0 $     &     0.27  &    0.325     &   15.3    & 17.6$^{\rm a}$  \\[0.2cm]
$^5$D$_0 \rightarrow \,^5$D$_1 $     &     2.4   &              &   3.1     & 2.34
\end{tabular}
\end{ruledtabular}
\begin{flushleft}
$^{\rm a}$The result of Ref.~\cite{Beck|2010|Improved_prediction} was adjusted to the present experimental energy of the $^5$D$_0 $ state.
\end{flushleft}
\end{table}

For the $^7$F$_J \rightarrow \,^7$F$_{J'}$ transitions, we observe very good agreement with the results reported in Ref.\,\cite{Beck|2010|Improved_prediction}. This agreement is expected, as these are allowed $M1$ transitions between fine-structure sublevels.
In the non-relativistic approximation, assuming the validity of $LS$ coupling, the reduced matrix elements of the M$1$ operator, ${\bm \mu} = - \mu_0 ({\bf J} + {\bf S})$ (where $\mu_0 = \alpha/2$ a.u. is the Bohr magneton), can be obtained analytically. Since samarium is not a particularly heavy atom, relativistic corrections to these matrix elements are expected to be relatively small.
\subsubsection{Parity violation amplitude}
\label{subsubsec:PV amplitude}
We carried out the calculation of the NSD PV amplitude for the
$4f^6 6s^2\, ^7$F$_0 \rightarrow \, 4f^6 6s^2\, ^5$D$_0$ transitions in $^{149}$Sm. The Hamiltonian describing 
the NSD PV electron-nuclear interaction is given by Eq.\,\eqref{Eq:NSD}.

We approximate the nucleus to be a uniformly charged ball: 
\begin{equation}
\rho({\bm r}) = \frac{3}{4 \pi R^3} \, \theta (R-r)\,,
\end{equation}
where $\theta (R-r)$ is the Heaviside step function. The function $\rho({\bf r})$ is normalized to a unity integral over the volume of the nucleus and is not intended to represent the {\it number density} of nucleons. Instead, it is introduced as a shape function describing the spatial distribution of the weak interaction inside the nucleus. With this convention, all information about the overall strength of the interaction is contained in the dimensionless constant $\kappa$.  After averaging over nuclear degrees of freedom, one obtains an effective electron-nucleus contact interaction proportional to the nuclear spin operator $\bm I$. The complicated nuclear physics is absorbed into $\kappa$.

The root-mean-square charge radii are $r_{\rm rms}$ = 4.9892\,fm and 5.0134\,fm for $^{147}$Sm and $^{149}$Sm, respectively \cite{Angeli|2013|nuclear_ground_state}, and, correspondingly, the nuclear radii for these two isotopes are
$R = \sqrt{5/3} \ r_{\rm rms}\approx 6.441$\,fm and $\approx 6.472$\,fm, respectively. 

If $|i \rangle$ and $|f \rangle$ are the initial and final atomic states of the same nominal parity, then, taking into account the NSD part of the PV interaction in the lowest nonvanishing order, one  can write  the electric dipole transition matrix element
as
\begin{widetext}
\begin{eqnarray}
\langle f | d_{q,\rm NSD}  | i \rangle  = \sum_n
\left[ \frac{\langle f | d_q | n  \rangle  \langle n | H_{\rm NSD} | i \rangle}{E_i - E_n}
+ \frac{\langle f | H_{\rm NSD} | n  \rangle \langle n | d_q | i \rangle}{E_f - E_n} \right],
\label{e2}
\end{eqnarray}
\end{widetext}
where $|... \rangle \equiv |J F m_{F} \rangle$, ${\bm F} = {\bm I} + {\bm J}$ is the total angular momentum, $m_F$ is the projection
of ${\bm F}$, 
$d_q$ is a component of the rank-one electric-dipole operator, 
and $H_{\rm NSD}$ is given by~\eref{Eq:NSD}.

The expression for the reduced matrix element (ME) of $d_{\rm NSD}$ was derived in~\cite{Porsev|2001|NSD_PNC_cal,Porsev|2012|Yb+_PV}: 
\begin{widetext}
\begin{eqnarray}
&& \langle J_f F_f || d_{\rm NSD} || J_i F_i \rangle  =  \sqrt{I(I+1)(2I+1)(2F_i+1)(2F_f+1)} \nonumber \\
&\times& \sum_{n} \left[ (-1)^{J_f - J_i}
\left\{ \begin{array}{ccc}
J_n  &  J_i  &   1    \\
 I   &   I   &  F_i   \\
\end{array} \right\}
\left\{ \begin{array}{ccc}
J_n  &  J_f  &  1   \\
F_f  &  F_i  &  I   \\
\end{array} \right\}
\frac{ \langle J_f || d || n, J_n \rangle \langle n, J_n || H_{\rm NSD} || J_i \rangle }{E_n - E_i} \right. \nonumber \\
&+& \left. (-1)^{F_f - F_i}
\left\{ \begin{array}{ccc}
 J_n  &  J_f  &   1    \\
  I   &   I   &  F_f   \\
\end{array} \right\}
\left\{ \begin{array}{ccc}
J_n  &  J_i  &  1   \\
F_i  &  F_f  &  I   \\
\end{array} \right\}
\frac{\langle J_f || H_{\rm NSD} ||n,J_n \rangle \langle n,J_n || d ||J_i \rangle}{E_n - E_f}  \right].
\label{Eq:dsd}
\end{eqnarray}
For the $^7$F$_0 \rightarrow \, ^5$D$_0$ transition ($F_i=F_f=I$) we obtain from \eref{Eq:dsd}
\begin{eqnarray}
&& \langle ^5\mathrm{D}_0\, I || d_{\rm NSD} || ^7\mathrm{F}_0\, I \rangle = \frac{1}{3}\sqrt{I(I+1)(2I+1)} \nonumber \\
&\times& \sum_n \left[ \frac{\langle ^5\mathrm{D}_0 || d || n,J_n=1 \rangle
                    \langle n,J_n=1 || H_{\rm NSD} || ^7\mathrm{F}_0 \rangle }{E_n - E_{^7\mathrm{F}_0}}
+ \frac{\langle ^5\mathrm{D}_0 || H_{\rm NSD} ||n,J_n=1 \rangle
        \langle n,J_n=1 || d || ^7\mathrm{F}_0 \rangle}{E_n - E_{^5\mathrm{D}_0}} \right]\,.
\label{5D7F}
\end{eqnarray}
Substituting $I=7/2$ for $^{147}$Sm or $^{149}$Sm in \eref{5D7F} and rearranging the MEs in the first term, we arrive at
\begin{eqnarray}
\langle ^5\mathrm{D}_0\, I || d_{\rm NSD} || ^7\mathrm{F}_0\, I \rangle = \sqrt{14}
\sum_n \left[ -\frac{\langle ^7\mathrm{F}_0 ||H_{\rm NSD}|| n \rangle \langle n || d || ^5\mathrm{D}_0 \rangle}{E_n - E_{^7\mathrm{F}_0}}
              + \frac{\langle ^5\mathrm{D}_0 ||H_{\rm NSD}||n \rangle \langle n || d || ^7\mathrm{F}_0 \rangle}{E_n - E_{^5\mathrm{D}_0}} \right].
\label{5D7F_1}
\end{eqnarray}
For subsequent calculations, it is convenient to write
\begin{eqnarray}
\langle ^5\mathrm{D}_0\, I || d_{\rm NSD} || ^7\!F_0\, I \rangle \equiv
  \langle ^7\mathrm{F}_0 || H_{\rm NSD} \cdot R_1 \cdot d || ^5\mathrm{D}_0 \rangle
+ \langle ^5\mathrm{D}_0 || H_{\rm NSD} \cdot R_2 \cdot d || ^7\mathrm{F}_0 \rangle\,,
\label{5D7F_2}
\end{eqnarray}
\end{widetext}
where we denote  the terms involving summations over $n$ by $R_1$ and $R_2$.

To calculate the NSD PV amplitude defined by Eq.\,\eqref{5D7F_1}, one needs to sum over all possible intermediate states or, alternatively, solve the corresponding inhomogeneous equation \cite{KozPorFla|1996|nuclear_anapole_moment}. 
In the present work, we solve the inhomogeneous equation, employing the Sternheimer–Dalgarno–Lewis method in the valence sector. We follow the approach suggested in~\cite{Sternheimer|1950|Nuclear_quadrupole,Dalgarno|1955|Long_range_force_cal} and previously applied by us to the calculation of PV amplitudes in Bi~\cite{KozPorFla|1996|nuclear_anapole_moment}, Fr~\cite{Porsev|2001|NSD_PNC_cal}, and Yb$^{+}$~\cite{Porsev|2012|Yb+_PV}. 

The results obtained in several approximations are summarized in~\tref{PNC}. Specifically, we performed calculations within the two-configuration, [$7s6pdf\!g$], [$9spdf$], and [$9spdf\!g$] configuration spaces. As seen from \tref{PNC}, the values of the PV amplitude vary significantly between these approximations. Since the transition involves different $^7\mathrm{F}_0$ and $^5\mathrm{D}_0$ states, the overall sign of the PV amplitude is not uniquely defined, 
therefore, we present its absolute value. The noticeable spread of the results indicates a substantial sensitivity to the choice of configuration space, suggesting that the present accuracy is limited.
\begin{table}
\caption{The nuclear spin-dependent PV amplitude (in units $i \kappa \cdot 10^{-14}\, |e| a_0$), where $a_0$ is
the Bohr radius. The values obtained in the two-configuration and [$7s6pdf\!g$], [$9spdf$], and [$9spdf\!g$] approximations
are listed in the respective columns. The total values are given in the last row in absolute value.
The present results do not include RPA corrections.}
\label{PNC}
\begin{ruledtabular}
\begin{tabular}{rccccc}
                                                                                    & two-conf. & [$7s6pdf\!g$] & [$9spdf$] & [$9spdf\!g$] \\
\hline \\[-0.7pc]
$\langle ^7\mathrm{F}_0 || H_{\rm NSD} \cdot R_1 \cdot d || ^5\mathrm{D}_0 \rangle$  &  0.62  &  -0.80        &  1.23     &  1.48      \\
$\langle ^5\mathrm{D}_0 || H_{\rm NSD} \cdot R_2 \cdot d || ^7\mathrm{F}_0 \rangle$  & -4.79  &   3.05        & -10.9     &  2.65      \\
$\langle ^5\mathrm{D}_0 \, I || d_{\rm NSD} || ^7\mathrm{F}_0\, I \rangle$           &  4.17  &   2.25        &  9.68     &  4.13
\end{tabular}
\end{ruledtabular}
\end{table}

For the [$9spdf\!g$] CI space we also carried out calculations including random-phase approximation (RPA) corrections to both the
electric-dipole and the NSD PV operators. In addition, when evaluating both contributions to
$\langle ^5$D$_0\, I || d_{\rm NSD} || ^7$F$_0 \, I \rangle$, the theoretical energies of the lowest-lying odd-parity intermediate states
were replaced with their experimental values.
This approach yields
\begin{equation}
\mathrm{E}1_{\rm PV} \equiv \langle ^5\mathrm{D}_0\, I || d_{\rm NSD} || ^7\mathrm{F}_0\, I \rangle \approx 3 \times 10^{-13}\, i \kappa \, |e| a_0\,,
\label{Eq:PNC_fin}
\end{equation}
which we consider to be our most comprehensive result and our best current estimate, accurate to within an order of magnitude.
We note that this value differs by approximately an order of magnitude from the result obtained without RPA corrections, underscoring the crucial role of these corrections in determining the PV amplitude. 

The initial and final states are many-electron states, and core excitations should also be taken into account.
However, in our previous study of the NSD PV amplitude in Yb$^+$~\cite{Porsev|2012|Yb+_PV}, the core contribution was found to be an order of magnitude smaller than the valence contribution. Given the overall limited accuracy of the present calculation, we neglect the core contribution at this stage.

\subsubsection{Hyperfine-interaction-induced $^7\!F_0 \rightarrow\, ^5\!D_0$ transition amplitudes and Stark-induced amplitude}\label{Sec:HFI_amplitudes}

While small on the scale of allowed M1 or E2 transitions, such amplitudes may still exceed the PV-induced amplitude by several orders of magnitude. The presence of such background amplitudes, in conjunction with experimental imperfections, may lead to systematics; thus, these amplitudes deserve attention when designing and executing an NSD APV experiment.

Of primary concern is the E2 amplitude that arises due to the hyperfine-interaction-induced mixing of the $^5$D$_0$ state with the $^7$P$_2$ state lying only 14.4\,cm$^{-1}$ below.
Our estimate yields for the reduced transition matrix element in the $J$ basis:
\begin{equation}
    \langle 6s^2\,^7\textrm{F}_0 || E2 || 6s5d\,^7\textrm{P}_2 \rangle  \approx 2\,|e|a_0^2 \,.
\end{equation}
With this, we evaluate the reduced HFS-induced amplitude for the transition of interest as
\begin{equation}
    \langle 6s^2\,^5\textrm{D}_0\, I || E2 || 6s^2\,^7\textrm{F}_0\,  I\rangle \approx 8 \times 10^{-6}\,( Q/{\rm b})\, |e|a_0^2
\label{E2hfs}    
\end{equation}

It is interesting to note that the two nonzero-$I$ isotopes of Sm have significantly different quadrupole moments $^{147}$Sm $Q$=-0.26(3)\,b and $^{149}$Sm $Q$=0.078(2)\,b, meaning that the HFS-induced transition amplitudes are also different, in the ratio of quadrupole moments ($\approx$-3.6). This will provide a valuable handle to control systematic effects in the PV experiment.

The HFS-induced E2 amplitude given by Eq.\,(\ref{E2hfs}) and E1$_{\rm PV}$ amplitude have different dimensions. 
A proper comparison of the E1 and E2 amplitudes requires the introduction of an additional dimensional factor, which can be obtained from the square root of the ratio of the corresponding E1 and E2 transition rates. We therefore define an effective E2 amplitude, E2$'$, as
\begin{eqnarray}
\mathrm{E2}' &=& \frac{\alpha \omega}{\sqrt{20}} \langle 6s^2\,^5\textrm{D}_0\, I \,|| \mathrm{E2} ||\, 6s^2\,^7\textrm{F}_0\,  I\rangle \nonumber \\
&\approx&   8 \times 10^{-10}\, (Q/{\rm b}) \,|e| a_0 ,
\end{eqnarray}
where $\omega$ is the transition frequency between the $^5\textrm{D}_0$ and $^7\textrm{F}_0$ states.

Then for the ratio we find
\begin{equation}
\frac{\mathrm{E1}_{\rm PV}}{\mathrm{E2}'} \sim 0.4 \times 10^{-3} \frac{\kappa}{Q/{\rm b}}.
 \end{equation}

There is also an HFS-induced M1 amplitude due to mixing with $J=1$ states of even parity. However, this amplitude is suppressed due to at least an order of magnitude larger energy separations for currently known level, as well as the additional suppression associated with the required spin flip and orbital angular momentum change.

In Cs, the M1 background amplitude was measured to be 
$\approx 4 \times10^{-5} \mu_0 
\approx 1.5\times10^{-7}\,\,{\rm a.u.}$ 
\cite{Bouchiat|1984|M1_background_PV}. 
The NSD PV amplitude in Cs 
is 5$\times 10^{-13}ie a_0 \kappa$ \cite{FlaMur97}.
The ratio of the NSD-PV amplitude to the HFS-induced E2$'$ background amplitude in the present case is more than two orders of magnitude larger for $^{147}$Sm than the ratio of the NSD-PV amplitude to the background amplitude in Cs. For $^{149}$Sm, there is a further suppression of E2$'$ by a factor of 3.6 due to the smallness of the nuclear quadrupole moment. This should significantly reduce systematic effects related to the background amplitude.

The main contribution to the corresponding Stark-induced amplitude comes from
\begin{equation*} 
A_{\mathrm{Stark}} \approx \frac{\mathcal{E}}{3} \sum_n 
 \frac{\langle {}^5D_0 || d || n \rangle \langle n || d || {}^7F_0 \rangle}{E_n - E_{^5\!D_0}},
\end{equation*}
where the sum runs over odd-parity intermediate states with $J=1$. The calculation was performed within the [$9spdf\!g$] CI method. To account for the complete set of intermediate states, we employed the method of solving an inhomogeneous equation. The most important contributions were improved by replacing the theoretical transition energies with their experimental values. We obtained $A_{\mathrm{Stark}} \approx 0.9\,\mathcal{E}$ (in a.u.), demonstrating the linear dependence of the induced amplitude on the applied electric field. For an electric field of $\mathcal{E} \sim 1\,\mathrm{kV/cm} \sim 2\times10^{-7}\,\mathrm{a.u.}$, this corresponds to $A_{\mathrm{Stark}} \sim 10^{-7}$\,a.u.
\subsection{Nuclear-spin-dependent PV-interaction constant}
\label{subsec:NSD-constant}

The dimensionless nuclear-spin-dependent interaction constant $\kappa$ may be represented as a sum
\begin{equation}
\kappa = \kappa_A + \kappa_{\mathrm{ax}} + \kappa_{\mathrm{hfs}}\,,
\label{eq:kappa_decomp_app}
\end{equation}
where $\kappa_A$ is the contribution of the nuclear anapole moment,
$\kappa_{\mathrm{ax}}$ is produced by the vector-electron--axial-nucleon
neutral-current interaction, and $\kappa_{\mathrm{hfs}}$ originates from the
combined action of the spin-independent weak interaction and the hyperfine
interaction.

Nuclear anapole moment usually gives the dominating contribution to $\kappa$ \cite{FlaKhr80}
.
In the single-particle valence-nucleon approximation, the anapole contribution
is given by \cite{FlaKhr80,FlaKhrSus84,FlaMur97}
\begin{align}
\kappa_A
&=
\frac{9}{10}\,\frac{\alpha\,\mu_i}{m_p r_0}\,g_i\,A^{2/3}\,\frac{K}{I+1}
\nonumber\\
&\approx
1.15\times 10^{-3}\,g_i\mu_i\,A^{2/3}\,\frac{K}{I+1}\,.
\label{eq:kappaA_app}
\end{align}
Here $\alpha\approx 1/137$ is the fine-structure constant, $m_p$ is the proton
mass, $r_0\approx 1.2~\mathrm{fm}$, $A$ is the mass number, and
$\mu_i$ is the magnetic moment of the valence nucleon in nuclear magnetons,
with $\mu_p\approx 2.793$ for the proton and $\mu_n\approx -1.913$ for the
neutron. The angular factor is
\begin{equation}
K=(I+\tfrac12)(-1)^{I-l_i+1/2}\,,
\label{eq:Kdef_app}
\end{equation}
where $l_i$ is the orbital angular momentum of the unpaired nucleon. The
constants $g_p$ and $g_n$ characterize the parity-violating nucleon--nucleus
potential. Following Refs.\,\cite{FlaKhrSus84,FlaMur97,Mansour,Fadeev}, for single-particle estimates one may use the values
\begin{equation}
g_p = 3.4\,, \qquad g_n = 0.9\,,
\label{eq:gpgn_app}
\end{equation}
obtained using the meson-exchange model. These numerical values of $g_p$ and $g_n$ correspond to the central values of the meson-nucleon interaction constants range \cite{Mansour,Fadeev}. The proton constant $g_p$ is dominated by the sum of $\rho$-meson and pion contributions, while for the neutron it is a difference of $\rho$ and $\pi$ contributions. The $ \rho$-proton PV interaction constant is accurately measured using the  PV effect in proton-proton scattering. Therefore, the measurement of $g_n$ is especially sensitive to the $\pi$ PV interaction constant, which is still poorly known. One could see here an analogy with the $g-2$ experiments on the measurements of lepton anomalous magnetic moments. 

Both  $^{147}$Sm and $^{149}$Sm nuclear ground states have negative parity and nuclear spin $I=7/2$.  Both anapole moments may therefore be estimated using Eq. (\ref{eq:kappaA_app}) with $I=7/2$, $l=3$ and $K=-4$, corresponding to a valence neutron in the orbital $f_{7/2}$. For $g_n=0.9$, we obtain $\kappa_A \approx 0.05$. However, the magnetic moments of these nuclei are different, -0.8148(7)\,$\mu_N$ and -0.6677(11)\,$\mu_N$ (the Schmidt model gives -1.913\,$\mu_N$). 
Moreover, Eq.\,(\ref{eq:kappaA_app}), like the Schmidt-model expression for the magnetic moment, was derived under the assumption of a single valence nucleon outside a zero-spin nuclear core producing a spherically symmetric potential. In contrast, $^{147}$Sm and $^{149}$Sm  have substantial electric quadrupole moments, -0.261(7)b and 0.075(8)b, respectively. Their opposite signs indicate deformations of opposite character and, consequently, substantial differences between their nuclear wave functions. The anapole moments of these two nuclei may therefore differ significantly.

The ratio of their anapole moments could be extracted with high accuracy from measurements of the ratio of APV effects in the two isotopes, because the uncertainties of the atomic calculations cancel in the isotope ratio.

The $Z^0$-boson-exchange contribution between the electron vector current and the
nucleon axial-vector current has the form \cite{FlaNovSusKhr1977,Ginges2004Review}
\begin{equation}
\kappa_{\mathrm{ax}} = C_2\,\frac{1/2-K}{I+1}\,,
\label{eq:kappaax_app}
\end{equation}
where $C_2\equiv -C_{2p}$ for a valence proton and
$C_2\equiv -C_{2n}$ for a valence neutron. The Standard Model couplings are
\begin{equation}
C_{2p}=-C_{2n}=\frac{g_A\bigl(1-4\sin^2\theta_W\bigr)}{2}
\approx 0.03\,,
\label{eq:C2_app}
\end{equation}
with $g_A\approx 1.26$ and $\sin^2\theta_W(0)\approx 0.2386$ in the convention used in
Refs.\,\cite{Ginges2004Review}.
This gives  $\kappa_{\mathrm{ax}} \approx 0.03$.

NSD PV interaction may also be induced by the combined action of NSI PV interaction and hyperfine interaction. Both hfs and weak interaction operators act in the vicinity of the nucleus. This allowed us to perform perturbation theory summation over intermediate states by solving the inhomogeneous Dirac equation in the nuclear Coulomb field, build an effective NSD PV operator and calculate the effect analytically presented as a hfs contribution to $\kappa$  \cite{FlaKhr1985}: 
\begin{equation}
\kappa_{\mathrm{hfs}}
=
-\frac{1}{3}\,Q_W\,\frac{\alpha\mu}{m_p r_0 A^{1/3}}
\approx
2.5\times 10^{-4}\,A^{2/3}\,\mu\,,
\label{eq:kappahfs_app}
\end{equation}
where $Q_W$ is the nuclear weak charge and $\mu$ is the nuclear magnetic moment
in nuclear magnetons.
The magnetic moments of $^{147}$Sm and $^{149}$Sm are  -0.8148(7)$\mu_N$ and -0.6677(11)$\mu_N$. This gives $\kappa_{\mathrm{hfs}} \approx -0.0057$ and $\kappa_{\mathrm{hfs}} \approx -0.0047$.

The constant  $\kappa_{\mathrm{hfs}}$ incorporates the contribution of high-energy single-particle electron excitations in the perturbation theory sum over intermediate states.  
 However, the Sm atom has a dense spectrum of levels, and this approach may underestimate the contribution of close levels mixed by the hfs interaction.
Therefore, this contribution from close levels should be estimated separately.
A noticeable contribution is produced by hfs mixing of the fine structure components.  
The first contribution to the hfs-induced PV transition between ground $^7$F$_0$ and excited $^5$D$_0$ states of the $4f^66s^2$ configuration is given by
\begin{equation}
 \langle ^7{\rm F}_0|d^{\mathrm{lower}}_{{\rm PV},z}|^5{\rm D}_0\rangle = 
 \frac{\langle ^7{\rm F}_0 |H_a| ^7{\rm F}_1\rangle\langle ^7{\rm F}_1|d^{\mathrm{lower}}_{{\rm PV},z}|^5{\rm D}_0\rangle}{E(^7{\rm F}_0)-E(^7{\rm F}_1)}\,,
\end{equation}
where $H_a$ is the magnetic-dipole HFS operator.
Using Eq.\,(7) from Ref.\,\cite{hfs}
\begin{eqnarray}
&& \langle JIF|H_a|J'IF\rangle = (-1)^{I+F+J'}\begin{Bmatrix} 
F & I & J\\ 
1 & J' & I 
\end{Bmatrix} \\
&\times& \sqrt{I(I+1)(2I+1)}g_I \langle J||V_a||J'\rangle\,, \nonumber
\end{eqnarray}
where $g_I =\mu/I$ and $V_a$ is electronic part of the magnetic dipole hfs operator,
and substituting $J=0$, $J'=1$, $F=I$ we get
\begin{equation}
 \langle JII|H_a|J'II\rangle = \sqrt{\frac{I+1}{3I}} \mu \langle J||V_a||J'\rangle\,.
 \end{equation}
 For $^{147}$Sm and $^{149}$Sm $I=7/2$. Then 
\begin{equation}
 \langle JII|H_a|J'II\rangle = \sqrt{\frac{3}{7}} \mu \langle J||V_a||J'\rangle
 \end{equation}
and
\begin{align}
 \langle ^7{\rm F}_0|&d^{\mathrm{lower}}_{{\rm PV},z}|^5{\rm D}_0\rangle  =\\ &\sqrt{\frac{3}{7}} \mu
 \frac{ \langle ^7{\rm F}_0||V_a||^7{\rm F}_1\rangle   \langle ^7\rm{F}_1|d^{\mathrm{lower}}_{{\rm PV},z}|^5{\rm D}_0\rangle}{E(^7{\rm F}_0)-E(^7{\rm F}_1)}\,,\nonumber 
\end{align}
which can also be written as 
\begin{equation}
 \langle ^7{\rm F}_0|d^{\mathrm{lower}}_{{\rm PV},z}|^5{\rm D}_0\rangle =   C \langle ^7{\rm F}_1|d^{\mathrm{lower}}_{{\rm PV},z}|^5{\rm D}_0\rangle\,,
\end{equation}
where $C$ is a dimensionless mixing coefficient
\begin{equation}
 C = \sqrt{\frac{3}{7}} \mu
 \frac{ \langle ^7{\rm F}_0||V_a||^7{\rm F}_1\rangle}{E(^7{\rm F}_0)-E(^7{\rm F}_1)}\,. 
\end{equation}
Numerical calculations give 
\begin{eqnarray}
&&\langle ^7{\rm F}_0||V_a||^7{\rm F}_1\rangle = 2.07 \times 10^{-7}  \ {\rm a.u.}, \\
&& \langle ^7{\rm F}_1|d^{\mathrm{lower}}_{{\rm PV},z}|^5{\rm D}_0\rangle = 1.06 \times 10^{-11} (-Q_W/N) iea_0.
\end{eqnarray}
For the energy denominator, we use the data from the NIST database~\cite{NIST_Dataset}
\begin{equation}
E(^7{\rm F}_0)-E(^7{\rm F}_1) 
= -1.333 \times 10^{-3} \ {\rm a.u.}.
\end{equation}
This leads to
\begin{equation}
C=-1.02 \times 10^{-4} \mu
\end{equation}
and the contribution to NSD PV amplitude 
\begin{equation}\label{hfsEPV} 
 \langle ^7{\rm F}_0|d^{\mathrm{lower}}_{{\rm PV},z}|^5{\rm D}_0\rangle =  - 1.08 \times 10^{-15} \mu (-Q_W/N) iea_0.
\end{equation}

There is also a much smaller contribution involving hfs mixing of the fine structure components in the upper state,
\begin{equation}
 \langle ^7{\rm F}_0|d^{\mathrm{upper}}_{{\rm PV},z}|^5{\rm D}_0\rangle = 
 \frac{\langle ^7{\rm F}_0 |d^{\mathrm{upper}}_{{\rm PV},z}| ^5{\rm D}_1\rangle\langle ^5{\rm D}_1|H_a|^5{\rm D}_0\rangle}{E(^5{\rm D}_0)-E(^5{\rm D}_1)}\,. 
\end{equation}
It can be written as
\begin{equation}
 \langle ^7{\rm F}_0|d^{\mathrm{upper}}_{{\rm PV},z}|^5{\rm D}_0\rangle = C' \langle ^7{\rm F}_0 |d^{\mathrm{upper}}_{{\rm PV},z}| ^5{\rm D}_1\rangle,
\end{equation}
with
\begin{equation}
 C' = \sqrt{\frac{3}{7}} \mu
 \frac{ \langle ^5{\rm D}_0||V_a||^5{\rm D}_1\rangle}{E(^5{\rm D}_0)-E(^5{\rm D}_1)}.
\end{equation}
Using calculated and NIST data, we get
\begin{eqnarray}
&&\langle ^5{\rm D}_0||V_a||^5{\rm D}_1\rangle = 4.81 \times 10^{-8} \ {\rm a.u.}, \\
&& \langle ^7{\rm F}_0|d^{\mathrm{upper}}_{{\rm PV},z}|^5{\rm D}_1\rangle = 9.30 \times 10^{-13} (-Q_W/N) iea_0, \\
&&E(^5{\rm D}_0)-E(^5{\rm D}_1)
= - 6.15 \times 10^{-3} \ {\rm a.u.}.
\end{eqnarray}
This leads to
\begin{equation}
C'=- 5.12 \times 10^{-6} \mu
\end{equation}
and
\begin{equation}\label{hfsEPVup}
 \langle ^7{\rm F}_0|d^{\mathrm{upper}}_{{\rm PV},z}|^5{\rm D}_0\rangle = - 4.76 \times 10^{-18} \mu (-Q_W/N) iea_0.
\end{equation}

The  total  NSD amplitude, dominated by the nuclear anapole moment and NSD $Z$ boson contributions, is  
\begin{equation}\label{Eq:PNC_fin_z}
\mathrm{E}1_{{\rm PV},z} \approx 2 \times 10^{-13}\, i \kappa \, |e| a_0\,,
\end{equation}
where $\kappa  \approx 0.08$.
Using $(-Q_W/N)=0.93$, we obtain that the contribution of the hfs mixing with close levels in the ground and excited states, Eqs.\,(\ref{hfsEPV},\ref{hfsEPVup}), gives effective $ \delta\kappa_{\mathrm{hfs}} \approx 0.004$, which is only slightly smaller than the contribution of the hfs mixing with high energy electron states, Eq.\,(\ref{eq:kappahfs_app}), which gives  $\kappa_{hfs} \approx -0.005$. Altogether, we see that the   PV hfs - induced amplitude is much smaller than the PV NSD amplitude produced by the nuclear anapole moment and NSD $Z$ boson contributions.

Our estimates have shown that the neutron skin effect is $\sim 0.3\%$ for NSI PV interaction in Sm.  However, as we have shown above, the contribution of NSI PV interaction is relatively small, so any uncertainty in the neutron skin contribution to the total PV amplitude is negligible.

Equations~(\ref{eq:kappa_decomp_app}) - (\ref{eq:kappahfs_app}) provide the standard
single-particle parametrization of the NSD interaction constant used to analyze
APV. For open-shell nuclei such as $^{147,149}$Sm, this
representation is best regarded as a convenient decomposition of the observable
constant $\kappa$ into anapole, axial-current, and hyperfine-induced parts; the
actual values of $\kappa_A$ and $\kappa_{\mathrm{ax}}$ can be substantially modified by nuclear deformation and many-body nuclear effects beyond the single-particle model.
Accurate determination of $\kappa_A$ and $\kappa_{\mathrm{ax}}$ requires dedicated nuclear many-body calculations, which are beyond the scope of the present paper.

\subsection{Laser-cooling scheme}
\label{Sec: Laser cooling}
To our knowledge, Sm atoms have not yet been laser cooled. This is largely due to the complexity of the spectrum (Fig.\,\ref{Fig:levels_full}) and the resulting uncertainty in branching ratios, arising from incomplete knowledge of the composition of the relevant states. Consequently, the development of an efficient cooling scheme is challenging, as no closed, fast optical cycling transitions appear to be available.
However, several strong transitions remain viable candidates for laser cooling, typically requiring $10^4$--$10^5$ scattering events to slow atoms from thermal speeds. (Here, we assume an atomic beam of Sm that is being decelerated in a Zeeman slower \cite{Metcalf|1999|Zeemanslower}.)  Their implementation would require multiple repump lasers to address population loss into other states. In particular, as mentioned above, the level at $22\,041.02~\mathrm{cm^{-1}}$ has the composition $58\%$ of $4f^66s6p\,\mathrm{{}^7F^\circ_0}$ and $32\%$ of $4f^55d6s^2\,$$\mathrm{{}^7F^\circ_0}$, while the remaining $10\%$ is unassigned \cite{NIST_Dataset}. The  $\mathrm{{}^7F_1 \rightarrow {}^7F^\circ_0}$ transition probability is $96(5)\times 10^6\,\mathrm{s^{-1}}$, as measured experimentally~\cite{Lawler2013}, corresponding to an upper bound on the upper-state lifetime of $\tau \approx 10~\mathrm{ns}$. This indicates that the transition is sufficiently fast for laser cooling, enabling an adequate cooling path length (less than 1\,m).
In the worst-case scenario for laser cooling, up to seven repump lasers may be required. However, this represents an upper limit, as the branching ratios are not known. The estimated decay rates from the ${}^7$F$_0^\circ$ state show that the combined contribution of other decay channels is only $\sim 0.1\%$. The level compositions indicate that strong overlap with one configuration reduces overlap with others, suggesting that the actual number of required repumps may be significantly smaller. To confirm this, experimental detection of the decay fluorescence is required. The corresponding transition wavelengths lie in spectral regions accessible with existing laser technology, supporting the feasibility of repumping and laser cooling of samarium. If multiple repump wavelengths are required, a Sm discharge source, such as a hollow-cathode lamp, could potentially provide multiple resonant wavelengths simultaneously. In addition, population trapped in dark states may be repumped through alternative allowed transitions originating from those states, which could further simplify the repumping scheme.
The energy of the $^7$D$_1$ state is currently unknown. Measurement of the fluorescence spectrum of the decays from the $^7$F$_0^\circ$ state would allow determination of the energy of the $^7$D$_1$ state and provide an experimental test of the theoretical calculations of the decay channels.

\bibliography{Samarium}

@article{Rochester1999Sm_lifetimes,
  author = {Rochester, S. and Bowers, C. J. and Budker, D. and DeMille, D. and Zolotorev, M.},
  title = {Measurement of lifetimes and tensor polarizabilities of odd-parity states of atomic samarium},
  journal = {Phys. Rev. A},
  publisher = {American Physical Society},
  volume = {59},
  issue = {5},
  pages = {3480--3494},
  year = {1999},
  month = {May},
  doi = {10.1103/PhysRevA.59.3480},
  numpages = {0},
}

@article{Lawler2013,
  author = {J E Lawler and A J Fittante and E A Den Hartog},
  title = {Atomic transition probabilities of neutral samarium},
  journal = {Journal of Physics B: Atomic, Molecular and Optical Physics},
  publisher = {{IOP} Publishing},
  volume = {46},
  number = {21},
  pages = {215004},
  year = {2013},
  month = {oct},
  doi = {10.1088/0953-4075/46/21/215004},
}

@article{Berengut2018,
  author = {Berengut, Julian C. and Budker, Dmitry and Delaunay, C\'edric and Flambaum, Victor V. and Frugiuele, Claudia and Fuchs, Elina and Grojean, Christophe and Harnik, Roni and Ozeri, Roee and Perez, Gilad and Soreq, Yotam},
  title = {Probing New Long-Range Interactions by Isotope Shift Spectroscopy},
  journal = {Phys. Rev. Lett.},
  publisher = {American Physical Society},
  volume = {120},
  issue = {9},
  pages = {091801},
  year = {2018},
  month = {Feb},
  doi = {10.1103/PhysRevLett.120.091801},
  numpages = {7},
}

@Book{Martin|1978|Energy_levels,
  author = {Martin, W.C. and Zalubas, Romuald and Hagan, Lucy},
  title = {Atomic Energy Levels - The Rare-Earth Elements: The Spectra of Lanthanum, Cerium, Praseodymium, Neodymium, Promethium, Samarium, Europium, Gadolinium, Terbium, Dysprosium, Holmium, Erbium, Thulium, Ytterbium, and Lutetium},
  publisher = {National Bureau of Standards},
  school = {National Bureau of Standards},
  year = {1978},
  doi = {10.6028/nbs.nsrds.60},
}

@techreport{Barkov1988,
  author = {Barkov, L. M. and Melik-Pashaev, D. A. and Zolotorev, M. S.},
  title = {Laser spectroscopy of atomic samarium},
  institution = {Institute of Nuclear Physics},
  number = {IYaF--88-142},
  pages = {52},
  year = {1988},
  address = {USSR},
  url = {http://inis.iaea.org/search/search.aspx?orig_q=RN:22029827},
}

@article{Pulhani2005,
  author = {Pulhani, A. and Shah, M. and Dev, Vas and Suri, Brij},
  title = {High-lying even-parity excited levels of atomic samarium},
  journal = {JOSA B},
  volume = {22},
  pages = {1117-1122},
  year = {2005},
  month = {05},
  doi = {10.1364/JOSAB.22.001117},
}

@article{Li2011,
  author = {Ming Li and Chang-Jian Dai and Jun Xie},
  title = {Even-parity states of the Sm atom with stepwise excitation},
  journal = {Chinese Physics B},
  publisher = {{IOP} Publishing},
  volume = {20},
  number = {6},
  pages = {063204},
  year = {2011},
  month = {jun},
  doi = {10.1088/1674-1056/20/6/063204},
}

@ARTICLE{Gomonai2012,
  author = {{Gomonai}, A.~I. and {Remeta}, E. Yu.},
  title = {{Investigation of highly excited states of samarium}},
  journal = {Optics and Spectroscopy},
  volume = {112},
  number = {1},
  pages = {15-23},
  year = {2012},
  month = {jan},
  doi = {10.1134/S0030400X12010092},
  adsurl = {https://ui.adsabs.harvard.edu/abs/2012OptSp.112...15G},
}

@ARTICLE{Shah2014,
  author = {{Shah}, M.~L. and {Sahoo}, A.~C. and {Pulhani}, A.~K. and
         {Gupta}, G.~P. and {Suri}, B.~M. and {Dev}, Vas},
  title = {{Investigations of high-lying even-parity energy levels of atomic samarium using simultaneous observation of two-color laser-induced fluorescence and photoionization signals}},
  journal = {European Physical Journal D},
  volume = {68},
  number = {8},
  pages = {235},
  eid = {235},
  year = {2014},
  month = {aug},
  doi = {10.1140/epjd/e2014-50076-8},
}

@article{Sahoo2020,
  author = {A.C. Sahoo and P.K. Mandal and M.L. Shah and Vas Dev},
  title = {Investigation of high-lying even-parity levels of atomic samarium with multi-color photoionization technique: Energies and radiative lifetimes},
  journal = {Journal of Quantitative Spectroscopy and Radiative Transfer},
  volume = {241},
  pages = {106714},
  year = {2020},
  doi = {10.1016/j.jqsrt.2019.106714},
  issn = {0022-4073},
}

@article{Feng2008,
  author = {Guan Feng and Dai Chang-Jian and Zhao Hong-Ying},
  title = {Studies on odd-parity states of the Sm atom},
  journal = {Chinese Physics B},
  publisher = {{IOP} Publishing},
  volume = {17},
  number = {10},
  pages = {3655--3661},
  year = {2008},
  month = {oct},
  doi = {10.1088/1674-1056/17/10/021},
  url = {https://doi.org/10.1088%2F1674-1056%2F17%2F10%2F021},
}

@article{LiMing2011,
  author = {Li, Ming and Dai, ChangJian and Xie, Jun},
  title = {A study on odd-parity high-lying states of the Sm atom with three-color resonant excitation},
  journal = {Science China: Physics, Mechanics and Astronomy},
  volume = {54},
  pages = {1124-1130},
  year = {2011},
  month = {06},
  doi = {10.1007/s11433-011-4263-7},
}

@article{Sufen1989,
  author = {Hu Sufen and Mei Shimin and Zhang Sen and Chen Xing and Xu Yunfei},
  title = {Observation and measurement of the autoionization spectra of atomic samarium},
  journal = {Chinese Physics Letters},
  publisher = {{IOP} Publishing},
  volume = {6},
  number = {2},
  pages = {64--67},
  year = {1989},
  month = {feb},
  doi = {10.1088/0256-307x/6/2/005},
}

@article{WenJie2009,
  author = {Qin Wen-Jie and Dai Chang-Jian and Xiao Ying and Zhao Hong-Ying},
  title = {Investigation of autoionization spectra of Sm atoms using an isolated-core excitation method},
  journal = {Chinese Physics B},
  publisher = {{IOP} Publishing},
  volume = {18},
  number = {5},
  pages = {1833--1837},
  year = {2009},
  month = {may},
  doi = {10.1088/1674-1056/18/5/019},
}

@article{Qin2010,
  author = {W.J. Qin and C.J. Dai and Y. Xiao},
  title = {The study of autoionizing states of the samarium atom},
  journal = {Journal of Quantitative Spectroscopy and Radiative Transfer},
  volume = {111},
  number = {7},
  pages = {997 - 1004},
  year = {2010},
  doi = {10.1016/j.jqsrt.2010.01.016},
  issn = {0022-4073},
}

@article{QinWenJie2010,
  author = {Wen-Jie Qin and Chang-Jian Dai and Ying Xiao},
  title = {Multistep excitation of autoionizing states of the samarium atom},
  journal = {Journal of Quantitative Spectroscopy and Radiative Transfer},
  volume = {111},
  number = {1},
  pages = {63 - 70},
  year = {2010},
  doi = {10.1016/j.jqsrt.2009.07.008},
  issn = {0022-4073},
}

@article{Jayasekharan2000,
  author = {T Jayasekharan and M A N Razvi and G L Bhale},
  title = {Even-parity bound and autoionizing Rydberg series of the samarium atom},
  journal = {Journal of Physics B: Atomic, Molecular and Optical Physics},
  publisher = {{IOP} Publishing},
  volume = {33},
  number = {16},
  pages = {3123--3136},
  year = {2000},
  month = {aug},
  doi = {10.1088/0953-4075/33/16/314},
}

@article{klaus2014,
  author = {Wendt, Klaus and Gottwald, Tina and Mattolat, C. and Raeder, Sebastian},
  title = {Ionization potentials of the lanthanides and actinides – towards atomic spectroscopy of super-heavy elements},
  journal = {Hyperfine Interactions},
  volume = {227},
  year = {2014},
  month = {06},
  doi = {10.1007/s10751-014-1041-8},
}

@article{Worden1978,
  author = {E. F. Worden and R. W. Solarz and J. A. Paisner and J. G. Conway},
  title = {First ionization potentials of lanthanides by laser spectroscopy},
  journal = {J. Opt. Soc. Am.},
  publisher = {OSA},
  volume = {68},
  number = {1},
  pages = {52--61},
  year = {1978},
  month = {Jan},
  doi = {10.1364/JOSA.68.000052},
}

@misc{NIST_Dataset,
  author = {Kramida, A. and Ralchenko, Yu. and Reader, J. and {NIST ASD Team}},
  title = {{NIST Atomic Spectra Database} (version 5.12)},
  year = {2024},
  doi = {10.18434/T4W30F},
  url = {https://physics.nist.gov/asd},
  howpublished = {National Institute of Standards and Technology, Gaithersburg, MD},
  note = {Online; accessed for Sm I level data},
}

@article{Dzuba1986,
  author = {Dzuba, VA and Flambaum, VV and Khriplovich, IB},
  title = {{Enhancement of P-and T-nonconserving effects in rare-earth atoms}},
  journal = {Zeitschrift f{\"u}r Physik D Atoms, Molecules and Clusters},
  publisher = {Springer},
  volume = {1},
  number = {3},
  pages = {243--245},
  year = {1986},
}

@article{Davies1989,
  author = {I O G Davies and P E G Baird and P G H Sandars and T D Wolfenden},
  title = {On the feasibility of detecting parity non-conserving optical rotation at 639 nm in {Sm I}},
  journal = {Journal of Physics B: Atomic, Molecular and Optical Physics},
  publisher = {{IOP} Publishing},
  volume = {22},
  number = {5},
  pages = {741--747},
  year = {1989},
  month = {mar},
  doi = {10.1088/0953-4075/22/5/006},
}

@article{Wolfenden1993,
  author = {T D Wolfenden and P E G Baird},
  title = {An experimental search for enhanced parity non-conserving optical rotation in samarium},
  journal = {Journal of Physics B: Atomic, Molecular and Optical Physics},
  publisher = {{IOP} Publishing},
  volume = {26},
  number = {7},
  pages = {1379--1387},
  year = {1993},
  month = {apr},
  doi = {10.1088/0953-4075/26/7/020},
}

@Article{Beck|2010|Improved_prediction,
  author = {Beck, Donald R and O’Malley, Steven M},
  title = {Improved {RCI} techniques for atomic {$4f^n$} excitation energies: application to {Sm I} {$4f{\hspace*{2pt}}^66s^{2}$} {$^5D_J$} levels},
  journal = {Journal of Physics B: Atomic, Molecular and Optical Physics},
  publisher = {IOP Publishing},
  volume = {43},
  number = {21},
  pages = {215003},
  year = {2010},
  month = {oct},
  doi = {10.1088/0953-4075/43/21/215003},
  issn = {1361-6455},
}

@article{Aramyan2025,
  author = {Aramyan, Razmik and Tretiak, Oleg and Sahoo, Sushree S. and Budker, Dmitry},
  title = {Enhanced multichannel dual-comb spectroscopy of complex systems},
  journal = {Phys. Rev. Appl.},
  publisher = {American Physical Society},
  volume = {24},
  issue = {2},
  pages = {L021002},
  year = {2025},
  month = {Aug},
  doi = {10.1103/7ktx-4h8m},
  url = {https://link.aps.org/doi/10.1103/7ktx-4h8m},
  numpages = {7},
}

@article{DzubaFlKozlov1996,
  author = {Dzuba, V. A. and Flambaum, V. V. and Kozlov, M. G.},
  title = {Combination of the many-body perturbation theory with the configuration-interaction method},
  journal = {Phys. Rev. A},
  publisher = {American Physical Society},
  volume = {54},
  issue = {5},
  pages = {3948--3959},
  year = {1996},
  month = {Nov},
  doi = {10.1103/PhysRevA.54.3948},
  url = {https://link.aps.org/doi/10.1103/PhysRevA.54.3948},
  numpages = {0},
}

@article{CIPT,
  author = {Dzuba, V. A. and Berengut, J. C. and Harabati, C. and Flambaum, V. V.},
  title = {Combining configuration interaction with perturbation theory for atoms with a large number of valence electrons},
  journal = {Phys. Rev. A},
  publisher = {American Physical Society},
  volume = {95},
  issue = {1},
  pages = {012503},
  year = {2017},
  month = {Jan},
  doi = {10.1103/PhysRevA.95.012503},
  url = {https://link.aps.org/doi/10.1103/PhysRevA.95.012503},
  numpages = {9},
}

@article{hfs,
  title = {{Hyperfine-induced transitions $^{1}$S$_{0}\text{\ensuremath{-}}^{3}$D$_{1}$ in Yb}},
  author = {Kozlov, M. G. and Dzuba, V. A. and Flambaum, V. V.},
  journal = {Phys. Rev. A},
  volume = {99},
  issue = {1},
  pages = {012516},
  numpages = {6},
  year = {2019},
  month = {Jan},
  publisher = {American Physical Society},
  doi = {10.1103/PhysRevA.99.012516},
  url = {https://link.aps.org/doi/10.1103/PhysRevA.99.012516}
}

@Article{FlaKhr80,
  author = {V V Flambaum and I B Khriplovich},
  title = {P-odd nuclear forces as a source of parity nonconservation in atoms},
  journal = {Zh. Eksp. Teor. Phys.},
  volume = {79},
  pages = {1656},
  year = {1980},
  note = {[Sov. Phys. JETP {\bf 52}, 835 (1980)]},
}

@Article{FlaKhrSus84,
  author = {V V Flambaum and I B Khriplovich and O P Sushkov},
  title = {Nuclear Anapole Moment},
  journal = {Phys. Lett. B},
  volume = {146},
  pages = {367},
  year = {1984},
}

@Article{FlaMur97,
  author = {Flambaum, V. V. and Murray, D. W.},
  title = {Anapole moment and nucleon weak interactions},
  journal = {Phys. Rev. C},
  volume = {56},
  pages = {1641--1644},
  year = {1997},
}

@article{FlaNovSusKhr1977,
  author = {Novikov, V. N. and Sushkov, O. P. and Flambaum, V. V. and Khriplovich, I. B.},
  title = {Possibility to investigate weak neutral current structure in optical transition of heavy atoms},
  journal = {Sov. Phys. JETP},
  volume = {46},
  pages = {420},
  year = {1977},
}

@article{FlaKhr1985,
  author = {Flambaum, V. V. and Khriplovich, I. B.},
  title = {New bounds  on the electric dipole moment of the electron and on T-odd electron-nucleon coupling},
  journal = {Sov. Phys. JETP},
  volume = {89},
  pages = {1505},
  year = {1985},
}

@article{WooBenCho97,
  author = {C. S. Wood and S. C. Bennett and  D. Cho and
                  B. P. Masterson and  J. L. Roberts and C. E. Tanner and  C. E. Wieman},
  title = {Measurement of parity nonconservation and an anapole moment in cesium},
  journal = {Science},
  volume = {275},
  number = {5307},
  pages = {1759--63},
  year = {1997},
  doi = {10.1126/science.275.5307.1759},
}

@article{Budker2002,
  author = {Budker, D. and Gawlik, W. and Kimball, D. F. and Rochester, S. M. and Yashchuk, V. V. and Weis, A.},
  title = {Resonant nonlinear magneto-optical effects in atoms},
  journal = {Rev. Mod. Phys.},
  publisher = {American Physical Society},
  volume = {74},
  issue = {4},
  pages = {1153--1201},
  year = {2002},
  month = {Nov},
  doi = {10.1103/RevModPhys.74.1153},
  url = {https://link.aps.org/doi/10.1103/RevModPhys.74.1153},
  numpages = {0},
}

@Article{Warrington|1995|PNC_Bismuth_Samarium,
  author = {Warrington, R B and Lucas, D M and Stacey, D N and Thompson, C D},
  title = {Atomic parity non-conservation: Recent measurements in bismuth and samarium},
  journal = {Physica Scripta},
  publisher = {IOP Publishing},
  volume = {T59},
  pages = {424--428},
  year = {1995},
  month = {Jan},
  doi = {10.1088/0031-8949/1995/t59/061},
  issn = {1402-4896},
}

@Article{Lucas|1998|PNC_opt_rotation,
  author = {Lucas, D. M. and Warrington, R. B. and Stacey, D. N. and Thompson, C. D.},
  title = {Search for parity nonconserving optical rotation in atomic samarium},
  journal = {Physical Review A},
  publisher = {American Physical Society (APS)},
  volume = {58},
  number = {5},
  pages = {3457--3471},
  year = {1998},
  month = {nov},
  doi = {10.1103/physreva.58.3457},
  issn = {1094-1622},
}

@article{barkov1989_Sm_levels_OS,
  author = {Barkov, L. M. and Zolotorev, M. S. and Melik-Pashaev, D. A.},
  title = {Study of the {$4f^6 6s^2\,{}^7F \rightarrow 4f^6 6s^2\,{}^5D$} forbidden transitions of atomic samarium},
  journal = {Optics and Spectroscopy (USSR)},
  volume = {66},
  pages = {288},
  year = {1989},
  note = {English translation of Opt. Spektrosk. 66, 495 (1989)},
}

@Article{Bouchiat|1975|PVpart2,
  author = {Bouchiat, M. A. and Bouchiat, C.},
  title = {{Parity violation induced by weak neutral currents in atomic physics. part II}},
  journal = {Journal de Physique},
  publisher = {EDP Sciences},
  volume = {36},
  number = {6},
  pages = {493--509},
  year = {1975},
  doi = {10.1051/jphys:01975003606049300},
  issn = {0302-0738},
}

@Article{Wood|1999|CJP_Cs_PNC,
  author = {Wood, C S and Bennett, S C and Roberts, J L and Cho, D and Wieman, C E},
  title = {Precision measurement of parity nonconservation in cesium},
  journal = {Canadian Journal of Physics},
  publisher = {Canadian Science Publishing},
  volume = {77},
  number = {1},
  pages = {7--75},
  year = {1999},
  month = {May},
  doi = {10.1139/p99-002},
  issn = {1208-6045},
}

@Article{Antypas|2018|APV_Yb,
  author = {Antypas, D. and Fabricant, A. and Stalnaker, J. E. and Tsigutkin, K. and Flambaum, V. V. and Budker, D.},
  title = {Isotopic variation of parity violation in atomic ytterbium},
  journal = {Nature Physics},
  publisher = {Springer Science and Business Media LLC},
  volume = {15},
  number = {2},
  pages = {120--123},
  year = {2018},
  month = {Oct},
  doi = {10.1038/s41567-018-0312-8},
  issn = {1745-2481},
}

@Article{Antypas|2019|PRA_APV_Yb,
  author = {Antypas, D. and Fabricant, A. M. and Stalnaker, J. E. and Tsigutkin, K. and Flambaum, V. V. and Budker, D.},
  title = {Isotopic variation of parity violation in atomic ytterbium: Description of the measurement method and analysis of systematic effects},
  journal = {Physical Review A},
  publisher = {American Physical Society (APS)},
  volume = {100},
  number = {1},
  pages = {012503},
  year = {2019},
  month = {July},
  doi = {10.1103/physreva.100.012503},
  issn = {2469-9934},
}

@Article{Childs|1972|HFS_Sm,
  author = {Childs, W. J. and Goodman, L. S.},
  title = {{Reanalysis of the hyperfine structure of the $4f^66s^2\,^7$F Multiplet in $^{147,149}$Sm, including measurements for the $^7$F$_6$ State}},
  journal = {Physical Review A},
  publisher = {American Physical Society (APS)},
  volume = {6},
  number = {6},
  pages = {2011--2021},
  year = {1972},
  month = {dec},
  doi = {10.1103/physreva.6.2011},
  issn = {0556-2791},
}

@book{Khr91,
  author = {I. B. Khriplovich},
  title = {Parity Nonconservation in Atomic Phenomena},
  publisher = {Gordon and Breach},
  year = {1991},
  address = {New York},
}

@Article{Ludlow|2015|RevModPhys_Optical_clocks,
  author = {Ludlow, Andrew D. and Boyd, Martin M. and Ye, Jun and Peik, E. and Schmidt, P.~O.},
  title = {Optical atomic clocks},
  journal = {Reviews of Modern Physics},
  publisher = {American Physical Society (APS)},
  volume = {87},
  number = {2},
  pages = {637--701},
  year = {2015},
  month = {June},
  doi = {10.1103/revmodphys.87.637},
  issn = {1539-0756},
}

@Article{Aeppli|2024|Sr_clock_10_19,
  author = {Aeppli, Alexander and Kim, Kyungtae and Warfield, William and Safronova, Marianna S. and Ye, Jun},
  title = {Clock with ${8\times10^{-19}}$ Systematic Uncertainty},
  journal = {Physical Review Letters},
  publisher = {American Physical Society (APS)},
  volume = {133},
  number = {2},
  pages = {023401},
  year = {2024},
  month = {jul},
  doi = {10.1103/physrevlett.133.023401},
  issn = {1079-7114},
}

@Article{Hinkley|2013|Yb_10_18,
  author = {Hinkley, N. and Sherman, J. A. and Phillips, N. B. and Schioppo, M. and Lemke, N. D. and Beloy, K. and Pizzocaro, M. and Oates, C. W. and Ludlow, A. D.},
  title = {An Atomic Clock with $10^{-18}$ Instability},
  journal = {Science},
  publisher = {American Association for the Advancement of Science (AAAS)},
  volume = {341},
  number = {6151},
  pages = {1215--1218},
  year = {2013},
  month = {sep},
  doi = {10.1126/science.1240420},
  issn = {1095-9203},
}

@Article{Dzuba|2018|4f_to_5d_Yb,
  author = {Dzuba, V. A. and Flambaum, V. V. and Schiller, S.},
  title = {Testing physics beyond the standard model through additional clock transitions in neutral ytterbium},
  journal = {Physical Review A},
  publisher = {American Physical Society (APS)},
  volume = {98},
  number = {2},
  pages = {022501},
  year = {2018},
  month = {aug},
  doi = {10.1103/physreva.98.022501},
  issn = {2469-9934},
}

@Article{Ishiyama|2023|Inner-Shell_clock_transition_Yb,
  author = {Ishiyama, Taiki and Ono, Koki and Takano, Tetsushi and Sunaga, Ayaki and Takahashi, Yoshiro},
  title = {Observation of an Inner-Shell Orbital Clock Transition in Neutral Ytterbium Atoms},
  journal = {Physical Review Letters},
  publisher = {American Physical Society (APS)},
  volume = {130},
  number = {15},
  pages = {153402},
  year = {2023},
  month = {apr},
  doi = {10.1103/physrevlett.130.153402},
  issn = {1079-7114},
}

@Article{Ishiyama|2026|Improvement_Yb_innershell,
  author = {Ishiyama, Taiki and Ono, Koki and Kawase, Hokuto and Takano, Tetsushi and Asano, Reiji and Sunaga, Ayaki and Yamamoto, Yasuhiro and Tanaka, Minoru and Takahashi, Yoshiro},
  title = {Orders-of-magnitude improvement in precision spectroscopy of an inner-shell orbital clock transition in neutral ytterbium},
  journal = {Nature Photonics},
  publisher = {Springer Science and Business Media LLC},
  year = {2026},
  month = {mar},
  doi = {10.1038/s41566-026-01857-8},
  issn = {1749-4893},
}

@article{Verma2019,
  author = {Verma, Mohit and Talamo, Luca and Sawaoka, Hiromitsu and Vutha, Amar C.},
  title = {Direct observation of a highly forbidden optical transition in {${\mathrm{Sm}:\mathrm{SrF}}_{2}$}},
  journal = {Phys. Rev. A},
  publisher = {American Physical Society},
  volume = {100},
  issue = {3},
  pages = {032502},
  year = {2019},
  month = {Sep},
  doi = {10.1103/PhysRevA.100.032502},
  url = {https://link.aps.org/doi/10.1103/PhysRevA.100.032502},
  numpages = {5},
}

@Article{Ahlefeldt|2016|Eu_first,
  author = {Ahlefeldt, R. L. and Hush, M. R. and Sellars, M. J.},
  title = {Ultranarrow Optical Inhomogeneous Linewidth in a Stoichiometric Rare-Earth Crystal},
  journal = {Physical Review Letters},
  publisher = {American Physical Society (APS)},
  volume = {117},
  number = {25},
  pages = {250504},
  year = {2016},
  month = {dec},
  doi = {10.1103/physrevlett.117.250504},
  issn = {1079-7114},
}

@Article{Sushkov|2023|Eu+3,
  author = {Sushkov, A. O. and Sushkov, O. P. and Yaresko, A.},
  title = {{Effective electric field: Quantifying the sensitivity of searches for new P,T -odd physics with $\mathrm{EuCl_3\cdot6H_2O}$}},
  journal = {Physical Review A},
  publisher = {American Physical Society (APS)},
  volume = {107},
  number = {6},
  pages = {062823},
  year = {2023},
  month = {jun},
  doi = {10.1103/physreva.107.062823},
  issn = {2469-9934},
}

@Book{Metcalf|1999|Zeemanslower,
  author = {Metcalf, Harold J. and van der Straten, Peter},
  title = {Laser Cooling and Trapping},
  journal = {Graduate Texts in Contemporary Physics},
  publisher = {Springer New York},
  year = {1999},
  doi = {10.1007/978-1-4612-1470-0},
  isbn = {9781461214700},
  issn = {0938-037X},
}

@Article{Door|2025|Ytterbium_IS_New_Bosons,
  author = {Door, Menno and Yeh, Chih-Han and Heinz, Matthias and Kirk, Fiona and Lyu, Chunhai and Miyagi, Takayuki and Berengut, Julian C. and Bieroń, Jacek and Blaum, Klaus and Dreissen, Laura S. and Eliseev, Sergey and Filianin, Pavel and Filzinger, Melina and Fuchs, Elina and Fürst, Henning A. and Gaigalas, Gediminas and Harman, Zoltán and Herkenhoff, Jost and Huntemann, Nils and Keitel, Christoph H. and Kromer, Kathrin and Lange, Daniel and Rischka, Alexander and Schweiger, Christoph and Schwenk, Achim and Shimizu, Noritaka and Mehlstäubler, Tanja E.},
  title = {Probing New Bosons and Nuclear Structure with Ytterbium Isotope Shifts},
  journal = {Physical Review Letters},
  publisher = {American Physical Society (APS)},
  volume = {134},
  number = {6},
  pages = {063002},
  year = {2025},
  month = {feb},
  doi = {10.1103/physrevlett.134.063002},
  issn = {1079-7114},
}

@article{Griffith1981,
  author = {J A R Griffith and G R Isaak and R New and M P Ralls},
  title = {Anomalies in the optical isotope shifts of samarium},
  journal = {Journal of Physics B: Atomic and Molecular Physics},
  publisher = {{IOP} Publishing},
  volume = {14},
  number = {16},
  pages = {2769--2780},
  year = {1981},
  month = {aug},
  doi = {10.1088/0022-3700/14/16/007},
  url = {https://doi.org/10.1088/0022-3700/14/16/007},
}

@Book{King|1984|IS_book,
  author = {King, W. H.},
  title = {Isotope Shifts in Atomic Spectra},
  publisher = {Springer US},
  year = {1984},
  doi = {10.1007/978-1-4899-1786-7},
  isbn = {9781489917867},
}

@article{Safronova2018,
  author = {Safronova, M. S. and Budker, D. and DeMille, D. and Kimball, Derek F. Jackson and Derevianko, A. and Clark, Charles W.},
  title = {Search for new physics with atoms and molecules},
  journal = {Rev. Mod. Phys.},
  publisher = {American Physical Society},
  volume = {90},
  issue = {2},
  pages = {025008},
  year = {2018},
  month = {Jun},
  doi = {10.1103/RevModPhys.90.025008},
  url = {https://link.aps.org/doi/10.1103/RevModPhys.90.025008},
  numpages = {106},
}

@inproceedings{Flambaum|2026|Alpha,
  author = {Flambaum, V. V.},
  title = {Atomic and Nuclear Clocks, Space-Time Variation of the Fundamental Constants and Dark Matter},
  booktitle = {Proceedings of the 1st International Online Conference on Atoms},
  publisher = {MDPI},
  year = {2026},
  month = {January 29--30},
  address = {Basel, Switzerland},
  note = {Published by MDPI/Sciforum},
}

@Article{Antypas|2020|Fast_oscil,
  author = {Antypas, Dionysios and Budker, Dmitry and Flambaum, Victor V. and Kozlov, Mikhail G. and Perez, Gilad and Ye, Jun},
  title = {Fast Apparent Oscillations of Fundamental Constants},
  journal = {Annalen der Physik},
  publisher = {Wiley},
  volume = {532},
  number = {4},
  year = {2020},
  month = {Mar},
  doi = {10.1002/andp.201900566},
  issn = {1521-3889},
}

@article{Stadnik|2015|Dark_matter_alpha,
  author = {Stadnik, Y. V. and Flambaum, V. V.},
  title = {Can Dark Matter Induce Cosmological Evolution of the Fundamental Constants of Nature?},
  journal = {Phys. Rev. Lett.},
  publisher = {American Physical Society},
  volume = {115},
  issue = {20},
  pages = {201301},
  year = {2015},
  month = {Nov},
  doi = {10.1103/PhysRevLett.115.201301},
  url = {https://link.aps.org/doi/10.1103/PhysRevLett.115.201301},
  numpages = {7},
}

@article{Tretiak2022_RFDMPRL,
  author = {Tretiak, Oleg and Zhang, Xue and Figueroa, Nataniel L. and Antypas, Dionysios and Brogna, Andrea and Banerjee, Abhishek and Perez, Gilad and Budker, Dmitry},
  title = {Improved Bounds on Ultralight Scalar Dark Matter in the Radio-Frequency Range},
  journal = {Phys. Rev. Lett.},
  publisher = {American Physical Society},
  volume = {129},
  issue = {3},
  pages = {031301},
  year = {2022},
  month = {Jul},
  doi = {10.1103/PhysRevLett.129.031301},
  url = {https://link.aps.org/doi/10.1103/PhysRevLett.129.031301},
  numpages = {7},
}

@article{Vedenin|1987|Pressure_broadening,
  author = {Vedenin, V D and Kulyasov, V N and Kurbatov, A L and Rodin, N V and Shubin, M V},
  title = {{$J$} dependence of collisional broadening for fine-structure components of {S}m {I}},
  journal = {Opt. Spectrosc. (Engl. Transl.); (United States)},
  volume = {62:4},
  year = {1987},
  month = {04},
  url = {https://www.osti.gov/biblio/5464592},
  issn = {ISSN OPSUA},
  place = {United States},
}

@article{Kra89,
  author = {Kraftmakher, A. {\rm Ya}},
  title = {Calculation of the amplitudes of {M1} and {E2} transitions between the {$4f^6 6s^2\,{}^7F$}  and {$4f^6 6s^2\,{}^5D$} states in the samarium atom},
  journal = {Opt. Spectrosc. (USSR)},
  volume = {66},
  pages = {565},
  year = {1989},
}

@article{Ferrara|2023|Spectral_lines,
  author = {Ferrara, C and Giarrusso, M and Leone, F},
  title = {Experimental atomic data of spectral lines - {I. Cs, Ba, Pr, Nd, Sm, Eu, Gd, Tb, Dy, Ho, Er, Tm, Yb, Lu, Hf, Re, and Os} in the 370-1000nm interval},
  journal = {Monthly Notices of the Royal Astronomical Society},
  publisher = {Oxford University Press (OUP)},
  volume = {527},
  number = {3},
  pages = {4440--4466},
  year = {2023},
  month = {oct},
  doi = {10.1093/mnras/stad3230},
  issn = {1365-2966},
}

@Book{Meggers|1975|Spectral_lines,
  author = {Meggers, William F and Corliss, Charles H and Scribner, Bourdon F},
  title = {Tables of spectral-line intensities: part 1- arranged by elements},
  publisher = {National Bureau of Standards},
  school = {National Bureau of Standards},
  year = {1975},
  doi = {10.6028/nbs.mono.145p1},
}

@article{Bogdanovich|1991|usage_transformed,
  author = {Bogdanovich, P},
  title = {Usage of transformed functions for calculations of electric dipole transitions},
  journal = {Lith. Phys. J.},
  volume = {31},
  pages = {79},
  year = {1991},
}

@article{KozPorFla|1996|nuclear_anapole_moment,
  author = {Kozlov, M. G. and Porsev, S. G. and Flambaum, V. V.},
  title = {{Manifestation of the nuclear anapole moment in the M1 transitions in bismuth}},
  journal = {J.  Phys.  B},
  volume = {29},
  pages = {689-97},
  year = {1996},
}

@Article{Angeli|2013|nuclear_ground_state,
  author = {Angeli, I. and Marinova, K. P.},
  title = {Table of experimental nuclear ground state charge radii: An update},
  journal = {Atomic Data and Nuclear Data Tables},
  publisher = {Elsevier BV},
  volume = {99},
  number = {1},
  pages = {69--95},
  year = {2013},
  month = {Jan},
  doi = {10.1016/j.adt.2011.12.006},
  issn = {0092-640X},
}

@Article{Porsev|2001|NSD_PNC_cal,
  author = {Porsev, S. G. and Kozlov, M. G.},
  title = {Calculation of the nuclear spin-dependent parity-nonconserving amplitude for the (7s,{F=}4)$\rightarrow$(7s,{F=}5) transition in {Fr}},
  journal = {Physical Review A},
  publisher = {American Physical Society (APS)},
  volume = {64},
  number = {6},
  pages = {064101},
  year = {2001},
  month = {Nov},
  doi = {10.1103/physreva.64.064101},
  issn = {1094-1622},
}

@Article{Porsev|2012|Yb+_PV,
  author = {Porsev, S. G. and Safronova, M. S. and Kozlov, M. G.},
  title = {Correlation effects in {Yb$^+$} and implications for parity violation},
  journal = {Physical Review A},
  publisher = {American Physical Society (APS)},
  volume = {86},
  number = {2},
  pages = {022504},
  year = {2012},
  month = {Aug},
  doi = {10.1103/physreva.86.022504},
  issn = {1094-1622},
}

@Article{Sternheimer|1950|Nuclear_quadrupole,
  author = {Sternheimer, R.},
  title = {On Nuclear Quadrupole Moments},
  journal = {Physical Review},
  publisher = {American Physical Society (APS)},
  volume = {80},
  number = {1},
  pages = {102--103},
  year = {1950},
  month = {Oct},
  doi = {10.1103/physrev.80.102.2},
  issn = {0031-899X},
}

@Article{Dalgarno|1955|Long_range_force_cal,
  author = {Dalgarno, Alexander and Lewis, J. T.},
  title = {The exact calculation of long-range forces between atoms by perturbation theory},
  journal = {Proceedings of the Royal Society of London. Series A. Mathematical and Physical Sciences},
  publisher = {The Royal Society},
  volume = {233},
  number = {1192},
  pages = {70--74},
  year = {1955},
  month = {Dec},
  doi = {10.1098/rspa.1955.0246},
  issn = {2053-9169},
}

@article{Ginges2004Review,
  author = {Ginges, J. S. M. and Flambaum, V. V.},
  title = {Violations of fundamental symmetries in atoms and tests of unification theories of elementary particles},
  journal = {Physics Reports},
  volume = {397},
  number = {2},
  pages = {63--154},
  year = {2004},
  doi = {10.1016/j.physrep.2004.03.005},
}

@Article{Berengut|2025|IS,
  author    = {Berengut, Julian C. and Delaunay, Cédric},
  journal   = {Nature Reviews Physics},
  title     = {Precision isotope-shift spectroscopy for new physics searches and nuclear insights},
  year      = {2025},
  issn      = {2522-5820},
  month     = Jan,
  number    = {2},
  pages     = {119--125},
  volume    = {7},
  doi       = {10.1038/s42254-024-00793-2},
  publisher = {Springer Science and Business Media LLC},
}

@Article{Bouchiat|1984|M1_background_PV,
  author    = {Bouchiat, M.A. and Guéna, J. and Pottier, L.},
  journal   = {Journal de Physique Lettres},
  title     = {{Measurement of the M1 amplitude and hyperfine mixing between the 6S$_{1/2}$-7S$_{1/2}$ caesium states}},
  year      = {1984},
  issn      = {0302-072X},
  number    = {2},
  pages     = {61--67},
  volume    = {45},
  doi       = {10.1051/jphyslet:0198400450206100},
  publisher = {EDP Sciences},
}

@article{Mansour,
  title = {Parity and time-reversal invariance violation in neutron-nucleus scattering},
  author = {Flambaum, V. V. and Mansour, A. J.},
  journal = {Phys. Rev. C},
  volume = {105},
  issue = {1},
  pages = {015501},
  numpages = {24},
  year = {2022},
  month = {Jan},
  publisher = {American Physical Society},
  doi = {10.1103/PhysRevC.105.015501},
  url = {https://link.aps.org/doi/10.1103/PhysRevC.105.015501}
}

@article{Fadeev,
  title = {Time-reversal invariance violation in neutron-nucleus scattering},
  author = {Fadeev, Pavel and Flambaum, Victor V.},
  journal = {Phys. Rev. C},
  volume = {100},
  issue = {1},
  pages = {015504},
  numpages = {7},
  year = {2019},
  month = {Jul},
  publisher = {American Physical Society},
  doi = {10.1103/PhysRevC.100.015504},
  url = {https://link.aps.org/doi/10.1103/PhysRevC.100.015504}
}

@article{CheKozPor25,
title = {p{CI}: A parallel configuration interaction software package for high-precision atomic structure calculations},
author = {C. Cheung and M. G. Kozlov and S. G. Porsev and M. S. Safronova and I. I. Tupitsyn and A. I. Bondarev},
journal= {Comput. Phys. Commun.},
volume = {308},
pages = {109463},
year = {2025},
doi = {10.1016/j.cpc.2024.109463}
}

@Article{Golovizin|2019|Tm_clock,
  author    = {Golovizin, A. and Fedorova, E. and Tregubov, D. and Sukachev, D. and Khabarova, K. and Sorokin, V. and Kolachevsky, N.},
  journal   = {Nature Communications},
  title     = {Inner-shell clock transition in atomic thulium with a small blackbody radiation shift},
  year      = {2019},
  issn      = {2041-1723},
  month     = Apr,
  number    = {1},
  volume    = {10},
  doi       = {10.1038/s41467-019-09706-9},
  publisher = {Springer Science and Business Media LLC},
}

@article{Beeks|2021|Th_clock,
  author  = {Beeks, Kjeld and Sikorsky, Tomas and Schumm, Thorsten
             and Thielking, Johannes and Okhapkin, Maxim V.
             and Peik, Ekkehard},
  title   = {The thorium-229 low-energy isomer and the nuclear clock},
  journal = {Nature Reviews Physics},
  volume  = {3},
  pages   = {238--248},
  year    = {2021},
  doi     = {10.1038/s42254-021-00286-6}
}

@article{Nima|2026|Limit_Schiff_moment_Eu,
  author  = {Bassam Nima and Mingyu Fan and Xubo Wang and Sen Wang and
             En Fu Zhou and Andrew M. Jayich and Jiang Ming Yao and
             Lan Cheng and Amar Vutha},
  title   = {{Limit on the nuclear Schiff moment of $^{153}$Eu}},
  journal = {arXiv},
  year    = {2026},
  eprint  = {2606.12084},
  url     = {https://arxiv.org/abs/2606.12084}
}

{\def\MakeTextUppercase#1{#1}\section*{ACKNOWLEDGMENTS}}
The authors acknowledge helpful discussions with M.G.\,Kozlov, Yu.\,Ralchenko, and K.\,Wendt, and the contributions to the experiment by A.\,Vutha, M.\,Smolis, and A.\,Posset.
This work has been supported by the Cluster of Excellence “Precision Physics,
Fundamental Interactions, and Structure of Matter” (PRISMA++ EXC 2118/2) funded
by the German Research Foundation (DFG) within the German Excellence Strategy
(Project ID 390831469) and by the Australian Research Council Grant No.\ DP230101058.
The theoretical work has been supported in part by the US NSF Award 2309254 and by the European Research Council (ERC) under the Horizon 2020 Research and Innovation Program of the European Union (Grant Agreement No. 856415).
{\def\MakeTextUppercase#1{#1}\section*{AUTHOR CONTRIBUTIONS}}
RA, KZ, and OT developed the experimental methods, conducted the experiments, processed the data, and wrote the manuscript. VAD, VVF, MSS, and SGP conducted many-body atomic calculations and cross-verified the results; SGP wrote parts of the manuscript. DB proposed and supervised the experiments and wrote the manuscript. All authors edited and proofread the paper.   
{\def\MakeTextUppercase#1{#1}\section*{COMPETING INTERESTS}}
The authors declare no competing interests.

\end{document}